\documentclass[prd,eqsecnum,twocolumn,amsfonts,amssymb]{revtex4}

\usepackage{graphicx}

\usepackage{bm}

\newcommand{\beq}{\begin{equation}}
\newcommand{\eeq}{\end{equation}}
\newcommand{\beqs}{\begin{eqnarray}}
\newcommand{\eeqs}{\end{eqnarray}}

\newcommand{\drawsquare}[2]{\hbox{%
\rule{#2pt}{#1pt}\hskip-#2pt
\rule{#1pt}{#2pt}\hskip-#1pt
\rule[#1pt]{#1pt}{#2pt}}\rule[#1pt]{#2pt}{#2pt}\hskip-#2pt
\rule{#2pt}{#1pt}}
\newcommand{\fund}{\raisebox{-.5pt}{\drawsquare{6.5}{0.4}}}

\begin{document}

\title{Scheme-Independent Calculations of Physical Quantities in an
${\cal N}=1$ Supersymmetric Gauge Theory} 

\author{Thomas A. Ryttov$^a$ and Robert Shrock$^b$}

\affiliation{(a) \ CP$^3$-Origins, University of Southern Denmark, Campusvej 
55, Odense, Denmark}

\affiliation{(b) \ C. N. Yang Institute for Theoretical Physics and 
Department of Physics and Astronomy \\
Stony Brook University, Stony Brook, NY 11794, USA }

\begin{abstract}

  We consider an asymptotically free, vectorial, ${\cal N}=1$ supersymmetric
  gauge theory with gauge group $G$ and $N_f$ pairs of chiral superfields in
  the respective representations ${\cal R}$ and $\bar {\cal R}$ of $G$, having
  an infrared fixed point (IRFP) of the renormalization group at
  $\alpha_{IR}$. We present exact results for the anomalous dimensions of
  various (gauge-invariant) composite chiral superfields $\gamma_{{\Phi}_{\rm
      prod}}$ at the IRFP and prove that these increase
  monotonically with decreasing $N_f$ in the non-Abelian Coulomb phase of the
  theory and that scheme-independent expansions for these anomalous dimensions
  as powers of an $N_f$-dependent variable, $\Delta_f$, exhibit monotonic and 
  rapid convergence to the exact $\gamma_{{\Phi}_{\rm prod}}$ throughout this 
  phase. We also present a
  scheme-independent calculation of the derivative of the beta function,
  $d\beta/d\alpha |_{\alpha=\alpha_{IR}}$, denoted $\beta'_{IR}$, up to
  $O(\Delta_f^3)$ for general $G$ and ${\cal R}$, and, for the case 
  $G={\rm SU}(N_c)$, ${\cal R}=F$, we give an analysis of the properties of
  $\beta'_{IR}$ calculated to  $O(\Delta_f^4)$. 

\end{abstract}

\maketitle


\section{Introduction}
\label{intro_section}

An important fact about quantum field theories is that their properties depend
on the Euclidean energy/momentum scale $\mu$ at which these properties are
measured. The change in these properties as a function of $\mu$ is described by
the renormalization group (RG). Asymptotically free gauge theories are
particularly amenable to renormalization-group analysis because the running
gauge coupling, $g(\mu)$, goes to zero in the limit of large $\mu$ in the deep
ultraviolet (UV), so that in this regime one can describe the theory accurately
using perturbative methods. The dependence of $g(\mu)$, or equivalently,
$\alpha(\mu) = g(\mu)^2/(4\pi)$, on $\mu$, is described by the beta function,
\beq
\beta = \frac{d\alpha}{dt} \ ,
\label{betadef}
\eeq
where $dt = d\ln\mu$.  

Here we consider an asymptotically free, vectorial, ${\cal N}=1$ supersymmetric
gauge theory with gauge group $G$ and $N_f$ pairs of massless chiral
superfields $\Phi_i$ and $\tilde \Phi_i$ transforming according to the
respective representations ${\cal R}$ and $\bar {\cal R}$ of $G$ \cite{fm}.  In
an asymptotically free theory of this type, as $\mu$ decreases from large
values in the UV toward $\mu=0$ in the infrared, $\alpha(\mu)$
increases. There are several possible types of infrared behavior, depending on
the gauge group and matter content of the theory.  We focus on the case in
which the beta function has a zero at a certain value $\alpha=\alpha_{IR}$,
which is an IR fixed point (IRFP) of the renormalization group.  Thus, as $\mu$
decreases from the UV to the IR, $\alpha(\mu)$ increases (monotonically) from 0
to the limiting value $\alpha_{IR}$.  In this IR limit, the theory is
scale-invariant, and is inferred to be conformally invariant \cite{scalecon}.
The combination of this conformal invariance with the supersymmetry means that
the theory is invariant under a superconformal algebra.  We denote the full
operator dimension of a physical (gauge-invariant) operator ${\cal O}$ as
$D_{\cal O}$.  In general, this can be written as
\beq
D_{\cal O} = D_{{\cal O},{\rm free}} - \gamma_{\cal O} \ , 
\label{anomdimdef}
\eeq
where $D_{{\cal O},{\rm free}}$ is the Maxwellian dimension that the operator
would have in a free theory and $\gamma_{\cal O}$ is the anomalous
dimension of ${\cal O}$ \cite{anomdimconv}. 

In this paper we present new scheme-independent results on the values of
physical quantities at this superconformal IR fixed point.  These quantities
include anomalous dimensions of gauge-invariant operators, $\gamma_{\cal O}$
and the derivative of the beta function, $\beta' \equiv d\beta/d\alpha$,
evaluated at $\alpha=\alpha_{IR}$ and thus denoted $\gamma_{{\cal O},IR}$ and
$\beta'_{IR}$. Specifically, we present exact results for anomalous dimensions
of various (gauge-invariant) composite chiral superfield operator products
$\Phi_{\rm prod}$ and study the properties of scheme-independent expansions of
these operators as power series in $\Delta_f$, where $\Delta_f$ is an
$N_f$-dependent expansion variable given in Eq. (\ref{delta}) below.
\cite{bz,gk,gg}. We prove that these anomalous dimensions increase
monotonically with decreasing $N_f$ in the non-Abelian Coulomb phase of the
theory and that scheme-independent expansions for these anomalous dimensions as
powers of $\Delta_f$ exhibit monotonic and rapid convergence to the exact
$\gamma_{{\Phi}_{\rm prod}}$ throughout this phase.  We also present a
scheme-independent calculation of $\beta'_{IR}$ up to $O(\Delta_f^3)$ for
general $G$ and ${\cal R}$ and analyze the properties of this expansion up to
$O(\Delta_f^4)$ for $G={\rm SU}(N_c)$ and ${\cal R}=F$, the fundamental
representation.  Previously, we have presented results for the anomalous
dimension $\gamma_{M,IR}$ of a meson-type chiral superfield using $n$-loop
series expansions and scheme-independent series expansions
\cite{bfs}-\cite{ir}. The current paper substantially extends our
earlier results.

This paper is organized as follows. Some relevant background and methods are
discussed in Section \ref{background_section}. In Section \ref{theorem_section}
we prove several theorems on anomalous dimensions of (gauge-invariant) chiral
superfields. In Sections \ref{gamma_m_section}-\ref{gamma_gen_section} we
present exact results on anomalous dimensions of various composite chiral
superfield operators.  These are generalized to theories with higher-dimension
matter chiral superfields in Section \ref{gamma_b_higherrep_section}. Section
\ref{betaprime_section} contains our results on $\beta'_{IR}$.  For the case
$G={\rm SU}(N_c)$ and ${\cal R}=F$, Section \ref{lnn_section} contains an
analysis of properties in the limit $N_c \to \infty$ and $N_f \to \infty$ with
the ratio $N_f/N_c$ fixed and finite. Our conclusions are given in Section
\ref{conclusion_section}.


\section{Background and Methods}
\label{background_section}

In this section we review some background and methods that we will use in our
calculations.  We consider an asymptotically free ${\cal N}=1$ supersymmetric
vectorial gauge theory with gauge group $G$ and $N_f$ copies (flavors) of
matter chiral superfields $\Phi^i$ and $\tilde \Phi_i$, $1 \le i \le N_f$,
transforming as the ${\cal R}$ and $\bar {\cal R}$ representations of $G$,
respectively.  We write the decomposition of the matter chiral superfield
$\Phi$ in terms of component fields (with group and flavor indices suppressed
here) as
\beq
\Phi = \phi + \sqrt{2} \, \theta \psi + \theta \theta F \ , 
\label{phisuperfield}
\eeq
where $\phi$, $\psi$, and $F$ are, respectively, the scalar, fermionic, and
auxiliary component fields, and $\theta$ is an anticommuting Grassmann
variable. The chiral superfield $W_\alpha$ contains the gluino $\lambda_\alpha$
and the field-strength tensor $F_{\mu\nu}^a$, where here $\alpha$ and $a$ are
spinor and gauge indices, respectively.  

The beta function of this theory has the series expansion 
\beq
\beta = -2\alpha \sum_{\ell=1}^\infty b_\ell \, \Big ( \frac{\alpha}{4\pi} \Big
)^\ell = -2 \alpha \sum_{\ell=1}^\infty \bar b_\ell \, \alpha^\ell \ , 
\label{beta}
\eeq
where $b_\ell$ is the $\ell$-loop coefficient and 
$\bar b_{\ell} = b_\ell/(4\pi)^\ell$. The first two coefficients, 
which are scheme-independent \cite{gross75}, are \cite{jones75,casimir}
\beq
b_1 = 3C_A - 2T_f N_f
\label{b1}
\eeq
and \cite{susyloops} 
\beq
b_2=6C_A^2-4(C_A+2C_f)T_fN_f \ .
\label{b2}
\eeq
The requirement of asymptotic freedom restricts $N_f$ to be less than 
an upper ($u$) bound $N_u$, i.e., 
\beq
N_f < N_u \ , 
\label{nfupper}
\eeq
where 
\beq
N_u = \frac{3C_A}{2T_f} \ . 
\label{Nfb1z}
\eeq
Note that $N_u$ is not necessarily an integer \cite{nfintegral}. 

The anomalous dimension of a (gauge-invariant) operator ${\cal O}$ has a series
expansion in powers of the coupling of the form 
\beq
\gamma_{\cal O} = \sum_{\ell=1}^\infty c_{{\cal O},\ell} \, 
\Big ( \frac{\alpha}{4\pi} \Big )^\ell \ , 
\label{gamma}
\eeq
where $c_{{\cal O},\ell}$ is the $\ell$-loop coefficient.  In particular, for a
chiral superfield $\Phi$, one may write 
\beq
\gamma_{\Phi} = \sum_{\ell=1}^\infty c_\ell \, 
\Big ( \frac{\alpha}{4\pi} \Big )^\ell \ . 
\label{gamma_series}
\eeq

From a calculation of the contribution of instantons to the action, 
Novikov, Shifman, Vainshtein, and Zakharov (NSVZ) derived a closed-form 
expression for the beta function \cite{nsvz}: 
\beq
\beta_{NSVZ} = -\frac{\alpha^2}{2\pi} \bigg [ \frac{b_1 -
  2N_fT_f \, \gamma_M}
{1-\frac{C_A\alpha}{2\pi}} \bigg ] \ , 
\label{beta_nsvz}
\eeq
where $\gamma_M$ is the anomalous dimension of the fermion bilinear that occurs
in the (gauge-invariant) quadratic chiral superfield operator product. We focus
here on the IR non-Abelian Coulomb phase (NACP), to be discussed further below,
in which the nonanomalous global chiral symmetry of the theory is exact.
Although we will analyze meson and baryon operators, as well as other
gauge-invariant products of chiral superfields later in the paper, it should be
kept in mind that there is no confinement in this NACP, and hence no physical
mesons or baryons.  The reason that we restrict to gauge-singlet operators is
so that the corresponding anomalous dimensions are gauge-invariant and hence
physical. 

In the NACP, a quadratic chiral superfield operator transforms according to an
(irreducible) representation of this global chiral symmetry.  Since the
anomalous dimensions are the same for these different representations (see,
e.g., \cite{gracey_gammatensor}), we denote the common anomalous dimension
simply as that for the singlet representation, corresponding to the quadratic
operator product $\tilde \Phi \Phi = \sum_{i=1}^{N_f} \tilde \Phi_i \Phi^i$.
Since this corresponds to the (gauge-invariant) fermion bilinear $\bar\psi\psi$
in a non-supersymmetric vectorial gauge theory, the anomalous dimension
$\gamma_M$ has often been denoted as $\gamma_{\bar\psi\psi}$ in our previous
papers \cite{bfs,gtr,gsi,dex,dexs,dexl}.

A number of exact results have been established about the (zero-temperature) IR
phase structure of the theory \cite{nsvz,seiberg,susyreviews}.  In the IR limit
$\mu \to 0$, $\alpha(\mu)$ approaches the limiting value $\alpha_{IR}$. In
particular, the theory flows from the UV to a non-Abelian Coulomb phase (NACP)
in the IR if 
\beq
{\rm NACP}: \quad N_\ell < N_f < N_u \ , 
\label{nacp}
\eeq
where 
\beq
N_\ell= \frac{3C_A}{4T_f} = \frac{N_u}{2} \ . 
\label{nell}
\eeq
As with $N_u$, note that $N_\ell$ is not necessarily an integer; it is the
actual physical lower end of the NACP if and only if it is an integer. In
particular, we note the important special case
\beqs
&& G={\rm SU}(N_c), \ \ {\cal R}=F \ \Longrightarrow \ 
N_\ell=\frac{3}{2}N_c, \ \ N_u=3N_c, \cr\cr
&&
\label{nacp_fund}
\eeqs
so that in this special case, $N_\ell$ is only physical if and only if $N_c$ is
even. This is to be understood implicitly below, when $N_\ell$ is referred to
as the lower end of the non-Abelian Coulomb phase
\cite{nellphysical}. Throughout the paper we will often consider a formal
generalization in which $N_f$ is analytically continued from the non-negative
integers to the (non-negative) real numbers, with the understanding that
physical values of $N_f$ are positive integers.  One reason for doing this is
to study the behavior of various quantities as $N_f$ approaches $N_u$ from
below and $N_\ell$ from above in the non-Abelian phase.

The two-loop beta function has an IR zero if $N_f$ is in the
interval $N_{f,b2z} < N_f < N_u$, where
\beq
N_{f,b2z} = \frac{3C_A^2}{2T_f(C_A+2C_f)} \ .  
\label{nfb2z}
\eeq
As we discussed in \cite{bfs}, $N_{f,b2z}$ may be larger than or smaller than
$N_\ell$, depending on the chiral superfield representation ${\cal R}$.  One
has
\beq
N_{f,b2z} - N_\ell = \frac{3C_A(C_A-2C_f)}{4T_f(C_A+2C_f)} \ . 
\label{ndif}
\eeq
This difference can be positive or negative. For the fundamental 
representation, ${\cal R}=F$, 
\beq
{\cal R}=F \ \Longrightarrow \ 
N_{f,b2z} - N_\ell = \frac{3N_c}{2(2N_c^2-1)} \ , 
\label{ndif_fund}
\eeq
which is positive.  However, for example, for the adjoint representation,
$R=adj$, this difference is negative: 
\beq
{\cal R}=adj: \ \Longrightarrow \ N_{f,b2z} - N_\ell = -\frac{1}{4} \ . 
\label{ndif_adj}
\eeq

For general $G$, the supersymmetric theory under consideration here is
invariant under a classical continuous global ($gb$) symmetry
\beqs
G_{cgb} &=& {\rm U}(N_f) \otimes {\rm U}(N_f) \otimes {\rm U}(1)_R \cr\cr
       &=& {\rm SU}(N_f) \otimes {\rm SU}(N_f) \otimes 
{\rm U}(1)_V \otimes {\rm U}(1)_A \otimes {\rm U}(1)_R \ , \cr\cr
&& 
\label{gcl}
\eeqs
where the first and second U($N_f$) groups consist of operators acting on 
$\Phi=(\Phi^1,...,\Phi^{N_f})$ and $\tilde\Phi=(\tilde\Phi_1,...,
\tilde\Phi_{N_f})$, respectively, and the U(1)$_R$ group is defined by the
commutation relations
\beq
[Q_\alpha,R]=Q_\alpha \ , \quad [Q_\alpha^\dagger,R]= -Q_\alpha^\dagger \ , 
\label{qrrel}
\eeq
where the $Q_\alpha$ and $Q_\alpha^\dagger$ are the 
generators of the supersymmetry transformations (with $\alpha$ a
spinor index here). The U(1)$_A$ symmetry is anomalous, due to
instantons, so the actual nonanomalous continuous global symmetry of the 
theory is 
\beqs
G_{gb} = {\rm SU}(N_f) \otimes {\rm SU}(N_f) \otimes {\rm U}(1)_V \otimes 
{\rm U}(1)_R \ . 
\label{gbl}
\eeqs
This symmetry is exact at a superconformal IRFP in the non-Abelian Coulomb
phase. Usually, for a U(1) (global or gauge) symmetry, the physics is invariant
under a multiplication of the charges of all fields by a nonzero real
constant. However, the situation is different for the U(1)$_R$
symmetry in a superconformal field theory; in this case, the 
$R$ charges of chiral superfields under the (global) U(1)$_R$ symmetry are 
uniquely determined \cite{susyreviews,mack,dim_rcharge_rel,susyreviews2,
intriligator_wecht}. 

The representations of the matter chiral superfields under the gauge and global
symmetry groups are listed in Table \ref{sgt_table} for the generic case in
which the representation ${\cal R}$ is complex. The case of (real) ${\cal R}$
will be discussed below.)  We recall the derivation of the $R$-charge
assignment to $\Phi$ and $\tilde \Phi$ (noting also that one can take
$R_\Phi=R_{\tilde \Phi}$). This assignment can be determined by the condition
that the U(1)$_R$ symmetry should not have a triangle anomaly determines the
$R$ charges of $\Phi$ (where the gauge and flavor indices are suppressed in the
notation).  The $R$ charge of the fermionic component $\psi$ in $\Phi$ is
$R_\psi = R_\Phi-1$. Given that $R_\lambda=1$ for the gluino, $\lambda$, the
sum of the contributions to the triangle anomaly from the gluino, and the
$\Phi$ and $\tilde \Phi$ matter superfields are $C_A + 2(R_\Phi-1)T_f N_f$. The
condition that this sum must be zero yields
\beq
R_\Phi = R_{\tilde \Phi} = 1 - \frac{C_A}{2T_f N_f} \ . 
\label{rcharge_phi}
\eeq
For the U(1)$_R$ symmetry to be non-anomalous, it is also necessary that,
similarly to the situation in non-supersymmetric theories, the one-loop
contribution is not modified by higher-order contributions, and this requisite
property holds \cite{susyanomaly}.

One can construct gauge-invariant quadratic operator products of the
``meson''-type, namely
\beq
M_i^j= \tilde \Phi_i \Phi^j \ , 
\label{meson}
\eeq
where, as above, $i$ and $j$ are flavor indices and the group indices are
implicit, with it being understood that they are contracted in such a way as 
to yield a singlet under the gauge group $G$.  As a
holomorphic product of chiral superfields, $M_i^j$ is again a chiral
superfield.  The fermionic bilinear operator product in $M_i^j$ is $\tilde
\psi_i \psi^j \equiv \tilde \psi_{i,L}^T C \psi^j_L$, where $C$ is the
conjugation Dirac matrix and we follow the usual convention of writing the
holomorphic chiral superfields as left-handed. Because the 
global symmetry (\ref{gbl}) is exact in the NACP, the meson-type quadratic 
chiral superfields transform according to (irreducible) representations of the
group $G_{gb}$.  We focus on the anomalous dimension of the diagonal operator
$\tilde\Phi \Phi = \sum_{i=1}^{N_f} \tilde \Phi_i \Phi^i$
evaluated at the IRFP $\alpha_{IR}$, which we denote as $\gamma_{M}$.

Consider next the case where $G={\rm SU}(N_c)$ and ${\cal R}=F$. The
transformation properties of the matter chiral superfields in this theory under
the global symmetry group $G_{gb}$ are listed in Table \ref{sqcd_table}. Since
we focus on the non-Abelian Coulomb NACP, where an IRFP is exact, $N_f$ must
lie in the interval $(3/2)N_c < N_f < 3N_c$.  Therefore, $N_f$ automatically
satisfies the requirement $N_f \ge N_c$ to construct the baryonic composite
chiral superfield operator
\beq
B^{i_1...i_{N_c}} = \epsilon_{a_1 \ldots a_{N_c}} 
\Phi^{a_1,i_1} \Phi^{a_2,i_2} \cdots \Phi^{a_{N_c},i_{N_c}}  
\label{baryon}
\eeq
and the corresponding operator involving the $\tilde \Phi$ chiral
superfields,
\beq
\tilde B_{i_1...i_{N_c}}=\epsilon_{a_1 \ldots a_{N_c}} \tilde \Phi_{i_1}^{a_1} 
\tilde \Phi_{i_2}^{a_2} \cdots \tilde \Phi_{i_{N_c}}^{a_{N_c}} \ , 
\label{tildebaryon}
\eeq
where here the $a_k$ and the $i_\ell$ are group and flavor indices,
respectively and $\epsilon_{a_1 \ldots a_{N_c}}$ is the totally antisymmetric
tensor density for the SU($N_c$) gauge group.  (If $N_f < N_c$, the operator
products (\ref{baryon}) and (\ref{tildebaryon}) vanish identically.).  Since
the flavors are equivalent with respect to the gauge interaction, we will
henceforth suppress the flavor dependence in the notation.  The full scaling
dimensions of $\Phi$ and $\tilde\Phi$ are equal, and the same is true for the
full scaling dimensions of $B$ and $\tilde B$, i.e., $D_{B,F} = D_{{\tilde
    B},F}$ (where the subscript indicates that ${\cal R}=F$, the fundamental
representation), so that the anomalous dimensions of these baryonic operators,
denoted $\gamma_{B,F}$ and $\gamma_{{\tilde B},F}$, are also equal. We thus
have
\beqs
D_{B,F} = D_{{\tilde B},F} &=& D_{B,F,{\rm free}} - \gamma_{B,F} \cr\cr
                   &=& N_c - \gamma_{B,F} \ . 
\label{dimb}
\eeqs
We shall discuss baryonic chiral superfield operator products for the
case where ${\cal R}$ is a higher-dimensional representation of $G$ later in
the paper.

In general (suppressing flavor indices), from $M$, $B$, and $\tilde B$, one can
construct a number of different composite gauge-invariant chiral
superfields. We denote such a generic composite chiral superfield consisting of
a (holomorphic) product of $n_M$ factors of a meson-type chiral superfield $M$
times $n_B$ factors of $B$ and $n_{\tilde B}$ factors of $\tilde B$ chiral 
superfields as $\Phi_{\rm prod}$:
\beq
\Phi_{\rm prod} = M^{n_M} B^{n_B} \tilde B^{n_{\tilde B}} \ . 
\label{phiprod}
\eeq
Here, to avoid cumbersome notation, the values of $n_M$, $n_B$, and $n_{\tilde
  B}$ are kept implicit in $\Phi_{\rm prod} \equiv \Phi_{{\rm prod};
  n_M,n_B,n_{\tilde B}}$.  One could also include a factor $(W_\alpha
W^\alpha)^{n_W}$, but (\ref{phiprod}) will be sufficient for our present
analysis.

There are several important quantities that characterize the properties of the
superconformal field theory at the IRFP at $\alpha_{IR}$. These include 
the derivative 
\beq
\beta'_{IR} \equiv \frac{d\beta}{d\alpha} \bigg |_{\alpha=\alpha_{IR}} 
\label{betaprime_def}
\eeq
and the anomalous dimensions of various gauge-invariant composite chiral
superfield operators evaluated at $\alpha=\alpha_{IR}$ such as $\gamma_M$,
$\gamma_B$, $\gamma_{{\tilde B}}$, and $\gamma_{\Phi_{\rm prod}}$. (Here and
below, we will often leave the dependence on ${\cal R}$ implicit in the
notation.)

As (gauge-invariant) physical quantities, $\beta'_{IR}$ and these
anomalous dimensions are
scheme-independent. However, the series expansions of these quantities in
powers of $\alpha$, calculated to a finite order, do not maintain this
scheme-independence beyond the lowest orders.  Hence, it is quite useful to
calculate and analyze series expansions for these quantities 
that are scheme-independent at each order. An important property is that
$\alpha_{IR} \searrow 0$ as $N_f \nearrow N_u$. This property is also shared by
a quantity that is manifestly scheme-independent, namely
\beq
\Delta_f = N_u-N_f \ , 
\label{delta}
\eeq
where $N_u$ was defined in Eq. (\ref{Nfb1z}).  
The maximal value of $\Delta_f$ in the NACP is
\beq
(\Delta_f)_{max,NACP} = N_u - N_\ell = \frac{N_u}{2} = \frac{3C_A}{4T_f} \ .  
\label{deltamax_nacp}
\eeq
As was observed by Banks and Zaks \cite{bz} (for a non-supersymmetric vectorial
gauge theory, in which $N_u = 11C_A/(4T_f)$), $\Delta_f$ is a natural
scheme-independent expansion variable.  In addition to \cite{bz}, some early
work with the $\Delta_f$ expansion was carried out in \cite{gk,gg}. In addition
to our previous works on scheme-independent series expansions 
\cite{gtr,gsi,dex,dexs,dexl}, see also \cite{kataev}. 

One may write a scheme-independent series expansions of $\beta'_{IR}$ in powers
of $\Delta_f$ as 
\beq
\beta'_{IR} = \sum_{j=2}^\infty d_j \, \Delta_f^j \ . 
\label{betaprime_ir_Deltaseries}
\eeq
In general, the calculation of $d_j$ requires, as inputs, the values of 
$b_\ell$ with $1 \le \ell \le j$. 

The property that $d_1=0$, so that $\beta'_{IR}$ vanishes like $\Delta_f^2$ as
$\Delta_f \to 0$, was derived in \cite{dex}. This property is general and does
not depend on whether the theory is supersymmetric or non-supersymmetric. A
simple way to understand this result is to note that for either type of theory,
the one-loop coefficient in the beta function has the form $b_1 =
b_{1,0}+b_{1,1}N_f$ (where $b_{1,0} > 0$ and $b_{1,1} < 0$), so that $N_u =
-b_{1,0}/b_{1,1}$.  Then, since $\Delta_f = N_u-N_f = -b_1/b_{1,1}$, it follows
that
\beq
\Delta_f \propto b_1 \ . 
\label{deltaf_propto_b1}
\eeq
From Eq. (\ref{betaprime}) below, $\beta'_{IR} = -2a_{IR}\sum_{\ell=1}^\infty
(\ell+1)b_\ell \, a_{IR}^{\ell-1}$, where 
$a_{IR}=\alpha_{IR}/(4\pi)$.  As $N_f \to N_u$, $\alpha_{IR}$ vanishes
linearly in $\Delta_f$, so in this limit, $\beta'_{IR} \propto \alpha_{IR}b_1
\propto \Delta_f^2$.

One may write the scheme-independent series expansion of $\gamma_M$ at the
superconformal IRFP in powers of $\Delta_f$ for a meson superfield operator: 
\beq
\gamma_M = \sum_{j=1}^\infty \kappa_j \, \Delta_f^j \ . 
\label{gamma_ir_Deltaseries}
\eeq
The calculation of $\kappa_j$ requires, as inputs, the values of $b_\ell$ with
$1 \le \ell \le j+1$ and $c_\ell$ with $1 \le \ell \le j$.  Similarly, the
scheme-independent series expansion of $\gamma_B = \gamma_{\tilde B, IR}$
at the IRFP in powers of $\Delta_f$ can be written as
\beq 
\gamma_B = \gamma_{\tilde B} = \sum_{j=2}^\infty f_{B,j}
\Delta_f^j \ .
\label{gamma_B_ir_Deltaseries}
\eeq
More generally, the scheme-independent expansion for a gauge-invariant
composite chiral superfield $\Phi_{\rm prod}$ consisting of a (holomorphic)
product of an arbitrary number of mesonic, baryonic, and conjugate baryonic
superfields, evaluated at the IRFP, can be written as
\beq
\gamma_{\Phi_{\rm prod}} = \sum_{j=2}^\infty f_{\Phi_{\rm prod},j} 
\Delta_f^j \ . 
\label{gamma_Phiprod_Deltaseries}
\eeq
These are thus series expansions extending downward below $N_u$ in the
non-Abelian Coulomb phase.  The truncations of these infinite
series to order $j=p$ inclusive are denoted 
$\beta'_{IR,\Delta_f^p} \equiv \beta'_{\Delta_f^p}$, $\gamma_{M,\Delta_f^p}$,
$\gamma_{B,\Delta_f^p}$, and 
$\gamma_{\Phi_{\rm prod},\Delta_f^p}$, respectively.

For a scalar operator (other than the identity), the condition of unitarity in
a conformal field theory implies the lower bound
\cite{mack,dim_rcharge_rel,gir} 
\beq
D_{\cal O} \ge 1 \ . 
\label{conformalbound}
\eeq
This bound holds regardless of whether the theory is supersymmetric or not.

In a supersymmetric conformal (i.e., superconformal) theory, one can take
advantage of additional information about the operator dimensions.
First, if a (composite or fundamental) chiral superfield ${\cal O}$ has 
$R$ charge $R_{\cal O}$, then \cite{susyreviews,mack,dim_rcharge_rel,
susyreviews2,gir,minwalla}
\beq
D_{\cal O} = \frac{3}{2} \, R_{\cal O} \ . 
\label{dim_rcharge_rel}
\eeq
We recall that since $D_{\cal O}$ is a physical quantity, the meaningfulness of
this relation depends on the fact that in a superconformal theory, the $R$
charges are uniquely determined. Since the U(1)$_R$ symmetry is exact in the
non-Abelian Coulomb phase considered here, the $R$ charge of an operator is a
conserved quantity. The $R$ charge of a holomorphic product of chiral
superfields is the sum of the $R$ charges of each of the chiral superfields in
the product:
\beq
R_{\Phi_{\rm prod}} = \sum_{k=1}^p R_{\Phi_k} \ . 
\label{rcharge_additivity}
\eeq
Hence, the full dimension of a holomorphic product $\Phi_{\rm
  prod}$ of chiral superfields $\Phi_k$, $k=1,...,p$, $\Phi_{\rm prod} =
\prod_{k=1}^p \Phi_k$, is the sum of the full dimensions of each chiral
superfield in the product (e.g., \cite{susyreviews2}):
\beq
D_{\Phi_{\rm prod}} = \sum_{k=1}^p D_{\Phi_k} \ . 
\label{d_additivity}
\eeq
Furthermore, the anomalous dimension of $\Phi_{\rm prod}$ is the sum of the
anomalous dimensions of the individual $\Phi_k$ superfields:
\beq
\gamma_{\Phi_{\rm prod}} = \sum_{k=1}^p \gamma_{\Phi_k} \ . 
\label{gamma_additivity}
\eeq
%


\section{Theorems on Properties of the Anomalous Dimensions of Composite Chiral
  Superfields }
\label{theorem_section}

In this section we prove some theorems on the properties of 
anomalous dimension $\gamma_{\Phi_{\rm prod}}$ of a gauge-invariant 
composite chiral superfield consisting of a (holomorphic) product of 
powers of $\Phi$ and/or $\tilde \Phi$ (where flavor indices are suppressed). 
Our results for the anomalous dimensions $\gamma_{\Phi_{\rm prod}}$ of various
particular composite chiral superfields given later in the paper will
illustrate these general theorems. 

The properties of the $R$ charge (\ref{rcharge_phi}) form the basis of the
resultant properties of the anomalous dimensions of the various composite
chiral superfields that we will consider.  We first use these properties to
prove a general monotonicity theorem concerning the anomalous dimension of a
chiral superfield operator containing products of $\Phi$ and/or
$\tilde\Phi$. This theorem applies for an arbitrary gauge group $G$ and fermion
representation.  We recall that $N_\ell = N_u/2$, as is evident in
Eqs. (\ref{Nfb1z}) and (\ref{nell}). For the following discussion, we
implicitly use the above-mentioned generalization of $N_f$ from non-negative
integers to real numbers.  As $N_f$ decreases from $N_u$ to $N_\ell$ in the
NACP, $R_\Phi$ decreases from 0 to $-1$.  Since the full scaling dimension of a
chiral superfield operator containing products of $\Phi$ and/or $\tilde \Phi$
satisfies (\ref{dim_rcharge_rel}) and since this full dimension is related to
the anomalous dimension of the operator according to (\ref{anomdimdef}), it
follows that the anomalous dimension $\gamma_{\cal O}$ is a monotonically
increasing function of decreasing $N_f$ in the NACP, which increases from
$\gamma_{\cal O}=0$ at the upper end of the NACP to a maximal value at the
lower end of the NACP.

We next prove a theorem on the structure the anomalous dimension 
of a general composite chiral superfield containing products of $\Phi$ and/or
$\tilde \Phi$, and the coefficients
$f_{\Phi_{\rm prod},j}$ in (\ref{gamma_Phiprod_Deltaseries}).  To do this, we
first express $R_\Phi = R_{\tilde \Phi}$ as a function of $\Delta_f$, obtaining
\beq
R_\Phi = 1 - \frac{1}{1-\frac{\Delta_f}{N_u}} \ . 
\label{rphi_delta_relation}
\eeq
Combining this with Eqs. (\ref{dim_rcharge_rel}) and (\ref{anomdimdef}), it
follows, as a second theorem, that the anomalous dimension of a general
composite chiral superfield containing products of $\Phi$ and/or $\tilde\Phi$,
evaluated at the superconformal IRFP, is of the form
\beqs
\gamma_{\Phi_{prod}} &=& C\Big [1-\frac{1}{1-\frac{\Delta_f}{N_u}} \Big ] 
\cr\cr 
&=& C\sum_{j=1}^\infty \Big ( \frac{\Delta_f}{N_u} \Big )^j \ , 
\label{gamma_structure}
\eeqs
where $C$ is a $\Delta_f$-independent constant depending on $G$, the fermion
representation, and the structure of $\Phi_{prod}$. Hence, as a corollary to
this theorem, we find that the coefficient $f_{\Phi_{\rm prod},j}$ of the
$O(\Delta_f^j)$ term in the expansion (\ref{gamma_Phiprod_Deltaseries}) is
given by
\beq
f_{\Phi_{\rm prod},j} = \frac{1}{N_u^j} \ . 
\label{fjgeneral}
\eeq
That is, up to an overall multiplicative factor $C$, $\gamma_{\Phi_{\rm prod}}$
is a geometric series in powers of $\Delta_f$, with the coefficients given in
Eq. (\ref{fjgeneral}).  As is evident in Eq. (\ref{fjgeneral}) is positive,
this coefficient $f_{\Phi_{\rm prod},j}$ is positive.  This leads to two
further monotonicity theorems.  Define $\gamma_{\Phi_{rm prod},\Delta_f^p}$ as
equal to the right-hand side of Eq.  (\ref{gamma_Phiprod_Deltaseries}) with the
upper limit $j=\infty$ replaced by $j=p$, i.e., the truncation of this infinite
series to order $O(\Delta_f^p)$.  Then the positivity of the coefficients
$f_{\Phi_{\rm prod},j}$ implies, as the third and fourth theorems, that (i) for
fixed $p$, the $O(\Delta_f^p)$ approximation, $\gamma_{\Phi_{rm
    prod},\Delta_f^p}$, to the exact $\gamma_{\Phi_{\rm prod}}$, is a
monotonically increasing function of $\Delta_f$, i.e., of decreasing $N_f$, and
(ii) for fixed $N_f$ and thus $\Delta_f$, $\gamma_{\Phi_{rm prod},\Delta_f^p}$
is a monotonically increasing function of the truncation order, $p$. We had
noted these monotonicity results in our earlier work for $\gamma_M$
\cite{gtr,gsi,dex,dexs,dexl}, and here we prove them in general.

A fifth theorem concerns the region of analyticity of the
expression for $\gamma_{\Phi_{prod}}$ in (\ref{gamma_structure}) and the
corresponding radius of convergence of the series expansion
(\ref{gamma_Phiprod_Deltaseries}) in powers of $\Delta_f$.  As is evident in
Eq. (\ref{gamma_structure}), this exact explicit expression for
$\gamma_{\Phi_{prod}}$ is an analytic function of $\Delta_f$ in the complex
$\Delta_f$ plane within a disk defined by
\beq
|\Delta_f| < N_u 
\label{deltaf_disk}
\eeq
and, correspondingly, the infinite series (\ref{gamma_Phiprod_Deltaseries})
converges for all $\Delta_f$ in this disk.  This region of convergence covers
the entire non-Abelian Coulomb phase because the maximal value of $\Delta_f$ in
this phase, as given by Eq. (\ref{deltamax_nacp}), is 
$(\Delta_f)_{max,NACP}=N_u/2$.


\section{Anomalous Dimension $\gamma_{M}$ }
\label{gamma_m_section}

In this section we discuss some results on $\gamma_{M}$ at a superconformal
IRFP that will be used in the paper. Since
\beq
R_M=R_\Phi+R_{\tilde \Phi} = 2\Big ( 1- \frac{C_A}{2T_fN_f} \Big ) \ , 
\label{r_meson}
\eeq
the full dimension of the quadratic chiral superfield operator $M$ (at the
superconformal IRFP) is 
\beqs
D_{M} &=& \frac{3}{2}R_M = 3\Big ( 1 - \frac{C_A}{2T_fN_f}\Big ) \cr\cr
    &=& 2-\gamma_{M} \ , 
\label{dim_meson}
\eeqs
and hence 
\beq
\gamma_{M} = \frac{3C_A}{2T_f N_f}-1  = \frac{N_u}{N_f} - 1 \ . 
\label{gamma_ir_meson_fund}
\eeq
where $N_u$ depends on ${\cal R}$. Expressing this anomalous dimension in
terms of $\Delta_f$, we have
\beq
\gamma_{M} = \frac{1}{1-\frac{\Delta_f}{N_u}}-1 = 
              \sum_{j=1}^\infty \bigg ( \frac{\Delta_f}{N_u} \bigg )^j \ , 
\label{gamma_ir_sgt}
\eeq
so the coefficient $\kappa_j$ in Eq. (\ref{gamma_ir_Deltaseries}) is
\beq
\kappa_j = \frac{1}{N_u^j} = \Big ( \frac{2T_f}{3C_A} \Big )^j \ . 
\label{kappajm_general}
\eeq
One sees that this general derivation is consistent with the NSVZ beta
function. This can be seen from the fact that at the IRFP, 
$\beta_{NSVZ}=0$; solving this equation yields the result
(\ref{gamma_ir_meson_fund}). Expressing $\gamma_{M}$ as a function of 
$\Delta_f$, we obtain the same results as in Eqs. (\ref{gamma_ir_sgt}) and
(\ref{kappajm_general}). 

For an ${\cal N}=1$ supersymmetric gauge theories with general $G$ and ${\cal
  R}$, $\gamma_{M}$ was calculated up to three-loop order in \cite{bfs} and
studied further in \cite{bc}-\cite{bfs2}. Concerning the scheme-independent
series expansion (\ref{gamma_ir_Deltaseries}), for general $G$ and ${\cal R}$,
$\kappa_1$ and $\kappa_2$ were calculated in \cite{gtr}, while for $G={\rm
  SU}(N_c)$ and ${\cal R}=F$, $\kappa_3$ was computed in \cite{dexs}. These
calculations used the beta function coefficients $b_1$-$b_4$ and 
the anomalous dimension coefficients $c_1$-$c_3$ from 
\cite{jones75,susyloops,susyloops2}.  Importantly, we
found that the results of our scheme-independent calculations of the $\kappa_j$
for this supersymmetric gauge theory agreed perfectly with the Taylor series
expansion of the exact expression (\ref{gamma_ir_sgt}).

Furthermore, as is evident from
the exact result (\ref{gamma_ir_sgt}), the small-$\Delta_f$ expansion of the
exact result is (absolutely) convergent for $|\Delta_f| < N_u$, i.e., 
\beq
|\Delta_f| < \frac{3C_A}{2T_f}  \ . 
\label{Delta_conv}
\eeq
This covers all of the non-Abelian Coulomb phase, which extends from
$N_u=3C_A/(2T_f)$ down to $N_\ell=N_u/2=3C_A/(4T_f)$, i.e., from $\Delta_f=0$
to $\Delta_f=3C_A/(4T_f)$. 

We next discuss the limiting values of $\gamma_M$ at a superconformal IRFP at
the upper and lower end of the NACP.  If one formally generalizes $N_f$ from
the positive integers to real numbers and lets $N_f$ decrease from $N_u$ to
$N_\ell$ in the NACP, $\gamma_{M}$ increases monotonically from 0 to 1,
saturating the upper bound allowed by conformal invariance at the lower end of
the NACP.  This behavior holds for general matter chiral superfield
representation $R$ and is a consequence of the fact that $N_\ell=N_u/2$.  As
stated, this is formal, because, in general, neither $N_u$ nor
$N_\ell$ is an integer, so the physical $N_f$, restricted as it is to integer
values, cannot necessarily take on either the value $N_u$ at which
$\gamma_{M}=0$ or the value $N_\ell$ at which $\gamma_{M} \nearrow 1$,
saturating the upper bound from conformality. In order for $N_f$ to be able to
reach $N_\ell$, it is necessary that $N_\ell$ be an integer.  In the case
$G={\rm SU}(N_c)$ with ${\cal R}=F$, (i) $N_u$ is always an integer, but (ii) 
since $N_\ell = (3/2)N_c$, it follows that
$N_\ell$ is an integer if and only if $N_c$ is even.  If, on the other hand,
$N_c$ is odd, then as $N_c$ decreases from $N_u=3N_c$ in the NACP, it cannot
actually reach $N_\ell$ since the latter is half-integral. In this case,
$\gamma_{M}$ does not saturate its conformality upper bound at the lower end of
the NACP. In this case where the matter chiral superfield representation is
${\cal R}=F$, one may avoid this complication by taking the limit $N_c \to
\infty$, $N_f \to \infty$ with the ratio $r=N_f/N_c$ fixed and finite.  As will
be discussed below, in this limit, $r$ is a real number and can always reach
the lower end of the non-Abelian Coulomb phase, so that $\gamma_{M}$ always
saturates its upper bound from conformal invariance.

It should be noted that the $\Delta_f$ expansion avoids a problem in which an
IRFP may not be manifest as a physical IR zero of the $n$-loop beta function
for some $n$.  Indeed, although the two-loop beta function, $\beta_{2\ell}$,
and the three-loop $\beta_{3\ell}$, calculated in the $\overline{DR}$ scheme,
have physical $\alpha_{IR,n\ell}$ zeros for $N_{f,b2z} < N_f < N_u$ in this
supersymmetric theory \cite{bfs}, we find that the four-loop beta function,
$\beta_{4\ell}$ (calculated in the $\overline{DR}$ scheme), does not exhibit a
physical IR zero, $\alpha_{IR,4\ell}$, for a substantial range of $N_f$ in this
interval. This is similar to what we found for $\alpha_{IR,5\ell}$ in the
non-supersymmetric gauge theory \cite{flir}.  In both cases, the $\Delta_f$
expansions (\ref{gamma_ir_Deltaseries}) and (\ref{betaprime_ir_Deltaseries})
circumvent this problem of a possible unphysical $\alpha_{IR,n\ell}$ that one
may encounter in using the conventional expansions (\ref{beta}).


\section{Anomalous Dimension $\gamma_{B}=\gamma_{{\tilde B}}$ for 
${\cal R}=F$}
\label{gamma_b_section}

In this section we specialize to the theory with gauge group $G={\rm SU}(N_c)$
and $N_f$ pairs of chiral superfields $\Phi^{a,i}$ and $\tilde \Phi_i^a$ (where
$a$ and $j$ are group and flavor indices) in the fundamental and conjugate
fundamental representations, denoted $F$ and $\bar F$, with Young tableaux
$\fund$ and $\overline{\fund}$, respectively. The matter content of this theory
is summarized in Table \ref{sqcd_table}

The $R$ charges of the basic chiral superfields are given in Table
\ref{sqcd_table}.  From Eq. (\ref{rcharge_additivity}), it follows that 
\beq
R_{B,F} = R_{{\tilde B},F} = N_c R_\Phi = 
N_c \Big (1 - \frac{N_c}{N_f} \Big ) \ . 
\label{r_baryon}
\eeq
Combining this with Eq. (\ref{dim_rcharge_rel}), one has the known exact 
result 
\beq
D_{B,F} = D_{{\tilde B},F} = 
\frac{3}{2} R_{B,F} = \frac{3}{2} N_c \Big ( 1 - \frac{N_c}{N_f} \Big ) \ . 
\label{dim_baryon}
\eeq
where we indicate ${\cal R}=F$ explicitly. Hence, the (equal) anomalous
dimensions of $B$ and $\tilde B$ at the superconformal IRFP are
\beq
\gamma_{B,F} = \gamma_{{\tilde B},F} = 
\frac{N_c}{2} \Big ( \frac{3N_c}{N_f} - 1 \Big ) \ . 
\label{gamma_ir_baryon}
\eeq
In Fig. \ref{baryon_Nc3_susy_fig} we plot the the value of $\gamma_{B,F}$ at 
the IRFP calculated to order $O(\Delta_f^p)$
with $1 \le p \le 3$, in comparison with the exact value,
Eq. (\ref{gamma_ir_baryon}), for the illustrative value $N_c = 3$. 
As was true of $\gamma_{M}$, we see that these $O(\Delta_f^p)$
truncations of the infinite series converge rapidly to the exact result.

Expressed as a function of $\Delta_f = 3N_c-N_f$, $\gamma_{B,F}$ is 
\beq
\gamma_{B,F} = \gamma_{{\tilde B},F} = \frac{N_c}{2} \, \bigg ( 
\frac{\frac{\Delta_f}{3N_c}}{1-\frac{\Delta_f}{3N_c}} \bigg ) \ . 
\label{gamma_ir_baryon_Deltaform}
\eeq
From Eqs. (\ref{gamma_ir_sgt}) and (\ref{gamma_ir_baryon}), one sees that 
$\gamma_{B,F}$ is simply proportional to $\gamma_{M,F}$: 
\beq
\gamma_{B,F} = \frac{N_c}{2} \, \gamma_{M,F} \ . 
\label{gamma_baryon_gamma_meson_relation}
\eeq

As $N_f \nearrow 3N_c$, i.e., $\Delta_f \searrow 0$, the common anomalous
dimension $\gamma_{B,F}=\gamma_{{\tilde B},F}$ vanishes, and as $N_f \searrow
(3/2)N_c$, i.e., $\Delta_f \nearrow (3/2)N_c$, it approaches the value
\beq
\lim_{N_f \searrow (3/2)N_c} \gamma_{(B, \tilde B)} = \frac{N_c}{2} 
\label{gamma_baryon_Nell}
\eeq
from below.

These baryonic composite chiral superfields have spin 0 (and are not equal to
the identity), so their respective full dimensions are bounded by the 
unitarity constraint from conformality, $D_B \ge 1$ and $D_{\tilde B}
\ge 1$.  This implies the upper bounds 
\beq
\gamma_{B,F} \le N_c - 1 \ , 
\label{gamma_baryon_bound}
\eeq
and thus also $\gamma_{{\tilde B},F} \le N_c - 1$.  Except for the case
$N_c=2$, where, owing to the reality of the representations of SU(2), the
baryonic and mesonic composite chiral superfield operators are equivalent, the
anomalous dimensions of the $B$ and $\tilde B$ operators at the IRFP do not
saturate their unitarity upper bound.  This is true, in particular, for the
infinite set of even values of $N_c$, for which $N_\ell$ is an integer and
hence is physical.  This behavior is in contrast to the situation that we found
for the anomalous dimension $\gamma_{M,F}$, which does saturate its upper
bound of 1 as $N_f \searrow N_\ell$ (assuming that $N_c$ is even so that
$N_\ell$ is an integer).


\section{Anomalous Dimensions of Composite Chiral Superfields} 
\label{gamma_gen_section}

In this section we derive exact expressions for the full dimension and hence
also the anomalous dimension of a variety of composite chiral superfields.  We
first discuss a SU($N_c$) theory with $N_f$ pairs of matter chiral superfields
$\Phi^i$ and $\tilde \Phi_i$, $i=1,...,N_f$, transforming as the $F$ and $\bar
F$ representations, respectively.  Our explicit results illustrate the general
theorems that we have proven above concerning these anomalous dimensions. 
We consider the composite chiral superfield $\Phi_{\rm prod}$ in
Eq. (\ref{phiprod}). Using Eqs. (\ref{rcharge_additivity}), we have
\beq
R_{\Phi_{\rm prod}} = \Big [ 2n_M + (n_B + n_{\tilde B})N_c \Big ]
 \Big (1- \frac{N_c}{N_f} \Big ) \ .
\label{rphiprod}
\eeq
Using Eq. (\ref{dim_rcharge_rel}), we have
\beq
D_{\Phi_{\rm prod}} = \frac{3}{2}\Big [2n_M + (n_B+n_{\tilde B})N_c \Big ]
\Big ( 1-\frac{N_c}{N_f} \Big ) \ . 
\label{dim_phiprod}
\eeq
Hence, 
\beq
\gamma_{\Phi_{\rm prod}}=
\Big [ n_M + \frac{(n_B+n_{\tilde B})}{2} \, N_c \Big ]
\Big ( \frac{3N_c}{N_f} -1 \Big )  \ . 
\label{gamma_phiprod}
\eeq
One sees that for the special case $(n_M,n_B,n_{\tilde B})=(1,0,0)$, 
the general result (\ref{gamma_phiprod}) reduces to Eq. 
(\ref{gamma_ir_meson_fund}), while for the special cases 
$(n_M,n_B,n_{\tilde B})=(0,1,0)$ and $(n_M,n_B,n_{\tilde B})=(0,0,1)$,
Eq. (\ref{gamma_phiprod}) reduces to Eq. (\ref{gamma_ir_baryon}). 
Expressing Eq. (\ref{gamma_phiprod}) as a function of $\Delta_f$ yields the
result
\beqs
\gamma_{\Phi_{\rm prod}}&=&
\Big [ n_M + \frac{(n_B+n_{\tilde B})}{2} \, N_c \Big ]
\bigg [ \frac{1}{1-\frac{\Delta_f}{3N_c}}-1 \bigg ] \cr\cr
&=& \Big [n_M + \frac{(n_B+n_{\tilde B})}{2} \, N_c \Big ] \, \sum_{j=1}^\infty
\Big ( \frac{\Delta_f}{3N_c} \Big )^j \ . \cr\cr
&&
\label{gamma_phiprod_Delta}
\eeqs
In agreement with our general monotonicity theorem proved above, this anomalous
dimension $\gamma_{\Phi_{\rm prod}}$ increases monotonically as a function of
$\Delta_f$ or equivalently decreasing $N_f$ in the NACP. As $N_f$ decreases
below $N_u$, $\gamma_{\Phi_{\rm prod}}$ increases monotonically from 0 to a
maximum of
\beq
\lim_{N_f \searrow (3/2)N_c} \gamma_{\Phi_{\rm prod}} = n_M + 
\Big ( \frac{n_B+n_{\tilde B}}{2} \Big ) N_c \ . 
\label{gamma_phiprod_Nell}
\eeq
From the conformality lower bound on the full dimension,
$D_{\Phi_{\rm prod}} \ge 1$, one obtains the corresponding upper bound
\beq
\gamma_{\Phi_{\rm prod}} \le 2n_M+(n_B+n_{\tilde B})N_c-1 \ .
\label{gamma_phiprod_bound}
\eeq

Expanding the exact expression in Taylor series, we read off the coefficient
$f_{\Phi_{\rm prod},j}$ as 
\beq
f_{\Phi_{\rm prod},j} = 
\Big [ n_M + \frac{(n_B+n_{\tilde B})}{2} \, N_c \Big ] \,
\Big ( \frac{1}{3N_c} \Big )^j \ . 
\label{fphiprodj}
\eeq
As is evident from Eq. (\ref{gamma_phiprod_Delta}), this series converges if
\beq
|\Delta_f| < 3N_c \ . 
\label{phiprod_convergence_criterion}
\eeq
This includes all of the NACP for this theory. 


\section{Baryonic Operators with Chiral Superfields in 
Higher-Dimensional Representations}
\label{gamma_b_higherrep_section}

\subsection{General} 

Here we derive corresponding exact results for anomalous dimensions of
(gauge-invariant) composite chiral superfield operators in (a vectorial,
asymptotically free, ${\cal N}=1$ supersymmetric) SU($N_c$) gauge theory
containing $N_f$ pairs of matter chiral superfields transforming according to
respective higher-dimensional representations ${\cal R}$ and $\bar {\cal R}$ of
the gauge group. As part of our analysis, we consider cases in which the
representation is real (or pseudoreal), i.e., ${\cal R} = \bar {\cal R}$.  For
a given type of higher-dimensional representation ${\cal R}$, the value of
$N_f$ is subject to the constraints that (i) the theory is asymptotically free,
so $N_f < N_u$, where $N_u$ was given in Eq. (\ref{Nfb1z}), and $N_f \ge
N_\ell$, where $N_\ell$ was given in Eq. (\ref{nell}), since we focus here on
an exact IRFP in the non-Abelian Coulomb phase.

For a general representation ${\cal R}$ of the matter chiral superfield, the
representations (charges) under the (anomaly-free) global symmetry can be read
from Table \ref{sgt_table}. If the gauge representation is
real or pseudoreal, then the global symmetry is enhanced, and the matter chiral
superfield has the representations given in Table
\ref{sgtreal_table}. Real representations include the (i) all representations
of SU(2), (ii) the adjoint representation of a general group $G$, and (iii) the
antisymmetric rank-$k$ representation of SU($2k$). 


\subsection{Adjoint Representation}

If ${\cal R}$ is the adjoint representation, then $N_u=3/2$ and $N_\ell=3/4$,
which allows just one Dirac value of $N_f$, namely $N_f=1$. Since the adjoint
representation is real, this is equivalent to $N_f=2$ Majorana chiral
superfields. Furthermore, owing to the reality of the adjoint representation,
composite superfields of baryon and meson type are equivalent. We denote these
by $M_i^j$, and they are written as
\beq
M_{ij} = \Phi^{a_1}_{\phantom{a_1}a_2,i} \Phi^{a_2}_{\phantom{a_2}a_1 j} =
\text{Tr}(\Phi_i \Phi_j) \ , 
\label{mesonij_adjoint}
\eeq
where the trace is over color indices, and $i,j$ are flavor indices. 
The full scaling dimension of this operator is
\beq
D_{M,adj} = \frac{3}{2} R_M = 3 \Big ( 1 - \frac{1}{2N_f} \Big ) \ , 
\label{dim_meson_adj}
\eeq
and therefore the anomalous dimension is
\beq
\gamma_{M,adj} = 2-3\Big ( 1 - \frac{1}{2N_f}\Big ) = \frac{3}{2N_f} -1 \ .
\label{gamma_meson_adj}
\eeq
Thus, $\gamma_{M,adj}$ takes the values 1/2 and $-1/4$ for the cases $N_f=1, \
2$, respectively. Note that these values are independent of $N_c$.  Expressed
as a function of $\Delta_f=N_u-N_f=(3/2)-N_f$, this anomalous dimension is
\beq
\gamma_{M,adj} = \frac{1}{1-\frac{2}{3}\Delta_f} -1 = \sum_{j=1}^\infty 
\Big (\frac{2\Delta_f}{3} \Big )^j \ . 
\label{gamma_meson_adj_series}
\eeq
We thus identify the coefficient $\kappa_j$ for this case as 
\beq
\kappa_{j,adj} =\Big ( \frac{2}{3} \Big )^j \ . 
\label{kappa_meson_adj}
\eeq
As before, formally continuing $N_f$ from its allowed integral values to real
values, we may study the properties of the small-$\Delta_f$ expansion to the
exact result. In Fig. \ref{adjoint_susy_fig} we plot $O(\Delta_f^p)$
approximations to $\gamma_{M,adj}$, together with the exact result. 
As is evident from this figure and from
Eq. (\ref{gamma_meson_adj}), finite truncations of this series converge rapidly
to the exact result in the NACP.  As we will see, this rapid convergence 
is also true of the other anomalous dimensions that we calculate below. 


\subsection{Rank-2 Symmetric Tensor Representation} 

Here we consider the case in which $G={\rm SU}(N_c)$ and ${\cal R}=S_2$, the
rank-2 symmetric tensor representation. If $N_c=2$, then the $S_2$
representation is the adjoint representation, which we have already discussed.
Therefore, we take $N_c \ge 3$.  Here,
\beq
N_{u,S_2} = \frac{3N_c}{N_c+2}
\label{Nu_sym}
\eeq
and
\beq
N_{\ell,S_2} = \frac{3N_c}{2(N_c+2)} \ , 
\label{Nell_sym}
\eeq
so that the non-Abelian Coulomb phase is comprised of the integer values of
$N_f$ in the formal interval $N_\ell \le N_f < N_u$, i.e., 
\beq
{\rm NACP}_{S_2}: \quad \frac{3N_c}{2(N_c+2)} \le N_f < \frac{3N_c}{N_c+2} \ . 
\label{nacp_sym}
\eeq
The condition that $N_f$ should be in the NACP restricts $N_f$.  For example,
for the values $N_c=3$ and $N_c=4$ the inequality 
(\ref{nacp_sym}) reads $9/10 < N_f < 9/5$ and $1 \le N_f < 2$, respectively,
allowing only the integer value $N_f=1$.  For $N_c =5$, the inequality
(\ref{nacp_asym}) reads $15/14 < N_f < 15/7$, allowing only the integer value
$N_f=2$, and more generally, for $N_c \ge 5$, the inequality (\ref{nacp_sym})
only allows the value $N_f=2$.  As $N_c \to \infty$, the inequality 
(\ref{nacp_sym}) approaches the limiting form $3/2 < N_f < 3$, with only the 
solution $N_f=2$.  

For $N_c \ge 3$, the $S_2$ representation is complex, so we consider both meson
and baryon chiral superfield operator products.  The meson product is
\beq
M_i^j = \tilde \Phi_{a_1 a_2,i} \Phi^{a_1a_2,j} = 
\text{Tr}(\tilde \Phi_i \Phi^j) \ , 
\label{meson_sym}
\eeq
where the trace is over the color indices and 
$\Phi^{a_1 a_2,i}=\Phi^{a_2 a_1,i} $. The full scaling dimension of this
operator is
\beq
D_{M,S_2} = \frac{3}{2} R_{M,S_2} = 
3 \Big [ 1 - \frac{N_c}{N_f(N_c+2)} \Big ] \ , 
\label{dim_meson_sym}
\eeq
and the anomalous dimension is
\beqs
\gamma_{M,S_2} &=& 2-3 \Big [ 1 - \frac{N_c}{N_f(N_c+2)} \Big ] \cr\cr
                  &=& \frac{3N_c}{N_f(N_c+2)} - 1 = \frac{N_{u,S_2}}{N_f}-1 
\ . 
\label{gamma_meson_sym}
\eeqs
As is clear from Eq. (\ref{gamma_meson_sym}), this is of the form 
(\ref{gamma_ir_meson_fund}) with $N_u=N_{n,S_2}$. 
Expressed in terms of $\Delta_f=N_u-N_f$, one obtains the special case of
(\ref{gamma_ir_sgt}) for the present theory with $N_u=N_{u,S_2}$ given by
(\ref{Nu_sym}).  As was the case with ${\cal R}=F$, since $N_\ell$ is not, in
general, an integer, $N_f$ cannot actually decrease all the way to be equal to
$N_\ell$, so $\gamma_{M,S_2}$ does not actually saturate its upper bound 
$\gamma_{M,S_2} \le 1$ from conformal invariance.  However, if 
one formally analytically continues $N_f$ from integers to 
real numbers, then this $N_f$ can decrease all the way to
$N_\ell$ at the lower boundary of the NACP, so $\gamma_{M,S_2}$ does
saturate this upper bound.  In Fig. \ref{meson_Nc3_2S_susy_fig} we plot 
$O(\Delta_f^p)$ approximations to $\gamma_{M,S_2}$, together with the exact 
result, for the case $N_c=3$.  We see again that finite truncations of this 
series converge rapidly to the exact result throughout the NACP. 

The baryon and antibaryon operators in this case are
\beq
B^{i_1,\ldots,i_{N_c}} = \frac{1}{N_c!} \epsilon_{a_1,\ldots,a_{N_c}}
\epsilon\
_{a'_1,\ldots,a'_{N_c}} \Phi^{a_1a'_1,i_1} \cdots \Phi^{a_{N_c}
  a'_{N_c},i_{N_c\
}}
\label{baryon_sym}
\eeq
and
\beq
\tilde{B}_{i_1,\ldots,i_{N_c}} = \frac{1}{N_c!}
\epsilon^{a_1,\ldots,a_{N_c}}\
 \epsilon^{a'_1,\ldots,a'_{N_c}} \tilde{\Phi}_{a_1a'_1,i_1} \cdots
 \tilde{\Phi}_{a_{N_c} a'_{N_c},i_{N_c}} \ . 
\label{antibaryon_sym}
\eeq
The way in which the color indices are contracted is similar to the 
determinant of a matrix. This is the reason we have included the 
$1/(N_c!)$ normalization factor. These operators have $R$ charge
\beq
R_{B,S_2} = R_{{\tilde B},S_2} =
 N_c \Big [ 1 - \frac{N_c}{N_f(N_c+2)} \Big ] \ . 
\label{r_baryon_sym}
\eeq
Hence, the full scaling dimensions of these operators are
\beqs
D_{B,S_2} = D_{{\tilde B},S_2} &=& \frac{3}{2}R_B \cr\cr
&=& \frac{3}{2}N_c \Big [ 1 - \frac{N_c}{N_f(N_c+2)} \Big ] \ . 
\label{dim_baryon_sym}
\eeqs
and the anomalous dimensions are
\beq
\gamma_{B,S_2} = \gamma_{{\tilde B},S_2} = 
\frac{N_c}{2} \Big [ \frac{3N_c}{N_f(N_c+2)} - 1 \Big ] \ . 
\label{gamma_baryon_sym}
\eeq
In Fig. \ref{baryon_Nc3_2S_susy_fig} we plot the $O(\Delta_f^p)$ 
approximations
to $\gamma_{B,S_2}$ for $G={\rm SU}(3)$, together with the exact result. 

The unitarity constraint for the baryons is the lower bound $D_B \ge 1$, and 
since $D_{B,S_2}=2N_c-\gamma_B$, this implies the upper bound
\beq
\gamma_{B,S_2} < N_c-1 \ . 
\label{gamma_baryon_sym_upperbound}
\eeq
Formally continuing $N_f$ to real numbers and evaluating $\gamma_B$ at
$N_f=N_\ell$, we find 
\beq
\gamma_{B,S_2} = \frac{N_c}{2} \quad {\rm at} \ N_f = N_\ell \ . 
\label{gamma_baryon_Nell_sym}
\eeq
For all $N_c \ge 3$, this does not saturate the upper bound
(\ref{gamma_baryon_sym_upperbound}). Furthermore, for most values of $N_c$, 
$N_\ell$ is not an integer, so the physical values of $N_f$ do not allow $N_f$
to actually decrease all the way to $N_\ell$, and hence the largest value of
$\gamma_{B,S_2}$ is actually smaller than $N_c/2$. 


\subsection{Rank-2 Antisymmetric Tensor Representation} 

We next consider the case in which $G={\rm SU}(N_c)$ and ${\cal R}=A_2$, the
rank-2 antisymmetric tensor representation. We restrict to $N_c \ge 4$, since
for $N_c=2$, then $A_2$ is the singlet and if $N_c=3$, then $A_2=\bar F$, the
conjugate fundamental. We have
\beq
N_{u,A_2} = \frac{3N_c}{N_c-2}
\label{Nu_asym}
\eeq
and
\beq
N_{\ell,A_2} = \frac{3N_c}{2(N_c-2)} \ , 
\label{Nell_asym}
\eeq
so that the non-Abelian Coulomb phase is comprised of the integer values of
$N_f$ in the formal interval $N_\ell \le N_f < N_u$, i.e., 
\beq
{\rm NACP}_{A_2}: \quad \frac{3N_c}{2(N_c-2)} \le N_f < \frac{3N_c}{N_c-2} \ .
\label{nacp_asym}
\eeq
As with the adjoint and $S_2$ representations, here also, the condition that
$N_f$ should be in the NACP restricts $N_f$.  For example, for the values
$N_c=4$ and $N_c=5$ the inequality (\ref{nacp_asym}) reads $3 \le N_f < 6$ and
$5/2 \le N_f < 5$, allowing only the integer values $N_f=3, 4$. For $N_f=8$,
the inequality (\ref{nacp_asym}) is $2 \le N_f < 4$, allowing only the values
$N_f=2, \ 3$.  As $N_c \to \infty$, the inequality (\ref{nacp_asym})
approaches the same limiting form as for $R=S_2$, namely, $3/2 < N_f < 3$, with
only the solution $N_f=2$. 

Here the meson-type chiral superfield product $M_i^j$ has the same form as
(\ref{meson_sym}), but with $\Phi^{a_1 a_2,i}=-\Phi^{a_2 a_1,i} $. The full 
scaling dimension of this operator is
\beq
D_{M,A_2} = \frac{3}{2} R_{M,A_2} = 
3 \Big [ 1 - \frac{N_c}{N_f(N_c-2)} \Big ] \ , 
\label{dim_meson_asym}
\eeq
and the anomalous dimension is
\beqs
\gamma_{M,A_2} &=& 2-3 \Big [ 1 - \frac{N_c}{N_f(N_c-2)} \Big ] \cr\cr
                  &=& \frac{3N_c}{N_f(N_c-2)} - 1 = \frac{N_{u,A_2}}{N_f}-1 \ .
\label{gamma_meson_asym}
\eeqs
Again, this is in accord with our general result 
(\ref{gamma_meson_sym}) with $N_u=N_{u,A_2}$, and again, this can be 
expressed as a function of $\Delta_f=N_u-N_f$, as in 
Eq. (\ref{gamma_ir_sgt}), with $N_u=N_{u,A_2}$. The same comments that were
made above apply here, namely that if one formally continues $N_f$ from the
integers to the real numbers, so that $N_f$ can decrease all the way to 
$N_\ell$, then $\gamma_{M,A_2}$ saturates its upper bound of 1.  However,
since $N_\ell$ is not, in general, an integer, so that $N_f$, restricted to
physical, integral values, cannot actually reach $N_\ell$, then, just as was
true with $\gamma_{M,S_2}$, $\gamma_{M,A_2}$ does not saturate its upper
bound from conformal invariance at the lower end of the NACP. 

In Figs. \ref{meson_Nc4_2A_susy_fig} and \ref{meson_Nc5_2A_susy_fig} we plot 
the anomalous dimension $\gamma_{M,A_2}$ to first, second
and third order in $\Delta_f$ for $N_c=4$ and $N_c=5$, together with the 
respective exact results. Note that for $N_c=4$ the $A_2$ representation is 
real, so the meson and baryon operators are equivalent.

For the baryons and antibaryons, we need to distinguish between even and odd
values of $N_c$. For even $N_c =2k $, these are
\beq
B^{i_1\cdots i_k} =  \frac{1}{2^k k!} \epsilon_{a_1,\ldots,a_{2k}}
\Phi^{a_1 a_2,i_1} \cdots \Phi^{a_{2k-1} a_{2k}, i_k}
\label{baryon_asym_Nceven}
\eeq
and
\beq
\tilde{B}_{i_1\cdots i_k} = \frac{1}{2^k k!}  \epsilon^{a_1,\ldots,a_{2k}}
\tilde{\Phi}_{a_1a_2,i_1} \cdots \tilde{\Phi}_{a_{2k-1} a_{2k}, i_k}
\label{antibaryon_asym_Nceven}
\eeq
while for odd $N_c$, they are
\beq
B^{i_1,\ldots,i_{N_c}} = \frac{1}{N_c!} \epsilon_{a_1,\ldots,a_{N_c}}
\epsilon_{a'_1,\ldots,a'_{N_c}} \Phi^{a_1a'_1,i_1} \cdots \Phi^{a_{N_c}
  a'_{N_c},i_{N_c}}
\label{baryon_asym_Ncodd}
\eeq
and
\beq
\tilde{B}_{i_1,\ldots,i_{N_c}} = \frac{1}{N_c!}
\epsilon^{a_1,\ldots,a_{N_c}} \epsilon^{a'_1,\ldots,a'_{N_c}} 
\tilde{\Phi}_{a_1a'_1,i_1} \cdots  \tilde{\Phi}\
_{a_{N_c} a'_{N_c},i_{N_c}}
\label{antibaryon_asym_Ncodd}
\eeq
Thus, for even and odd values of $N_c$, the respective baryon operators 
involves $N_c/2=k$ and $N_c$ $A_2$ chiral superfields. Correspondingly, 
for even and odd $N_c$, the contractions of the color indices are analogous 
to a Pfaffian and a determinant, respectively. 

For even $N_c$ (denoted $Nce$), the full scaling dimension of the baryon and
antibaryon operators is
\beq
D_{B,A_2,Nce} = D_{\tilde{B},A_2,Nce} = 
\frac{3N_c}{4} \Big [ 1 - \frac{N_c}{N_f(N_c-2)}\Big ]
 \ , 
\label{dim_baryon_asym_Nceven}
\eeq
so the anomalous dimension is
\beq
\gamma_{B,A_2,Nce} = \gamma_{\tilde{B},A_2,Nce} = \frac{N_c}{4} 
\Big [ \frac{3N_c}{N_f(N_c-2)} - 1 \Big ] \ . 
\label{gamma_baryon_asym_Nceven}
\eeq
We plot $\gamma_{B,A_2,Nce}$ for $N_c=6$ in Fig. \ref{baryon_Nc6_2A_susy_fig}.

The unitarity constraint from conformal invariance is again $D_B > 1$, and 
since $D_B = (N_c/2) - \gamma_B$, this implies the upper bound
\beq
\gamma_{B,A_2,Nce} < \frac{N_c}{2} -1 \ .
\label{gamma_baryon_upperbound_asym_Nceven}
\eeq
If one formally analytically continues $N_f$ to the real numbers, as
discussed above, so that $N_f$ can decrease all the way to $N_\ell$ in the
NACP, then the maximal value of $\gamma_{B,A_2}$ is
\beq
\gamma_{B,A_2,Nce} = \frac{N_c}{4} \quad {\rm at} \ N_f = N_{\ell,A_2} \ . 
\label{gamma_baryon_asym_Nceven_Nell}
\eeq
If $N_c=4$, then at $N_\ell=1$, $\gamma_{B,A_2,Nce}$ reaches a maximum value
of 1, saturating the unitarity upper bound $\gamma_{B,A_2,Nce} \le 1$ from
conformal invariance.  For even $N_c \ge 6$, the maximum value of
$\gamma_{B,A_2,Nce}$ as $N_f$ formally decreases to $N_\ell$ does not
saturate the unitarity upper bound, since $N_c/4 < (N_c/2)-1$ for $N_c \ge
6$. As $N_c \to \infty$ through even values, the ratio of the maximum value of
$\gamma_{B,A_2,Nce}$ evaluated at the formal (non-integral) value of
$N_\ell$ divided by the unitarity upper bound from conformal invariance
approaches 1/2.

For odd $N_c$ (denoted $Nco$), the full scaling dimension of the baryon is
\beq
D_{B,A_2,Nco} = D_{\tilde{B},A_2,Nco} = 
\frac{3N_c}{2} \Big [ 1 - \frac{N_c}{N_f(N_c-2)} \Big ] \ , 
\label{dim_baryon_asym_Ncodd}
\eeq
so the corresponding anomalous dimension is
\beq
\gamma_{B,A_2,Nco} = \gamma_{\tilde{B},A_2,Nco} = 
\frac{N_c}{2} \Big [ \frac{3N_c}{N_f(N_c-2)} - 1 \Big ] \ .
\label{gamma_baryon_asym_Ncodd}
\eeq
We plot $\gamma_B$ for $N_c=5$ in Fig. \ref{baryon_Nc5_2A_susy_fig}.

The unitarity constraint from conformal invariance is again $D_{B,A_2} \ge 1$,
and since $D_{B,A_2,Nco} = (N_c/2)-\gamma_{B,A_2,Nco}$, this implies the upper
bound
\beq 
\gamma_{B,A_2,Nco} < N_c -1 \ .
\label{gamma_baryon_upperbound_asym_Ncodd}
\eeq
With the same analytic continuation as above, 
\beq
\gamma_{N,A_2,Nco} = \frac{N_c}{2} \quad {\rm at} \ N_f=N_{\ell,A_2} \ . 
\label{gamma_baryon_asym_Ncodd_Nell}
\eeq
Even with an analytic continuation of $N_f$ from the integers to the real
numbers so that $N_f$ can actually reach down to $N_{\ell,A_2}$, this 
never saturates the unitarity upper bound from
conformal invariance at the lower end of the NACP, since 
$(N_c/2) < N_c-1$ for $N_c \ge 3$. 


\section{Scheme-Independent Calculation and Analysis of $\beta'_{IR}$}
\label{betaprime_section}

\subsection{General} 

In this section we study the scheme-independent expansion for the derivative of
the beta function evaluated at the superconformal IR fixed point, denoted
$\beta'_{IR}$, in the non-Abelian Coulomb phase.  Specifically, we present our
calculations of the scheme-independent coefficients $d_2$ and $d_3$ for general
$G$ and ${\cal R}$ and analyze the properties of $d_4$ and $\beta'_{IR}$
calculated to $O(\Delta_f^4)$ for the case $G={\rm SU}(N_c)$ and ${\cal
  R}=F$. For this special case $G={\rm SU}(N_c)$ and ${\cal R}=F$, quantities
equivalent to the $d_j$ were calculated in \cite{gg} for $2 \le j \le 4$. Our
new contributions here are calculations of $d_2$ and $d_3$ for general $G$ and
${\cal R}$ and also a different analysis of $\beta'_{IR}$ in the lower part of
the non-Abelian Coulomb phase.  One of the reasons for interest in this
derivative is that $\beta'_{IR}$ is equivalent \cite{grisaru97} to the
anomalous dimension of the Konishi supercurrent \cite{konishi}.


\subsection{Calculation via Series Expansion in $\alpha$} 

It is useful first to review the calculation of $\beta'_{IR}$ in \cite{bc,lnn}
using a conventional series expansion in powers of $\alpha$ up to three-loop
order.  In general, from Eq. (\ref{beta}), it follows that 
\beq
\beta' = -2\sum_{\ell=1}^\infty (\ell+1) b_\ell \, a^\ell \ . 
\label{betaprime}
\eeq
where $a = \alpha/(4\pi) = g^2/(16\pi^2)$. 
Evaluating the $n$-loop truncation of this series at the IR zero in the
$n$-loop beta function, $\alpha_{IR,n\ell}$ yields the $n$-loop value 
of the derivative, $\beta'_{IR,n\ell}$. 
Since $b_1$ and $b_2$ are scheme-independent, this is also true of 
$\beta'_{IR,2\ell}$, for which one finds \cite{bc} 
\beqs
\beta'_{IR,2\ell} &=& - \frac{2b_1^2}{b_2} \cr\cr
&=& \frac{(3C_A-2T_fN_f)^2}{2(C_A+2C_f)T_fN_f-3C_A^2} \ . 
\label{betaprime_2loop}
\eeqs
For general $G$ and ${\cal R}$, $\beta'_{IR,2\ell}$ increases monotonically as
$N_f$ decreases from $N_u$ in the NACP.  At the three-loop level, the condition
for the IR zero is the quadratic equation $b_1 + b_2 a + b_3 a^2=0$, whence,
$a^2=-(b_1+b_2a)/b_3$. Substituting this into Eq. (\ref{betaprime}), one has
\beq
\beta'_{IR,3\ell} = 2a_{IR,3\ell}(2b_1 + b_2 a_{IR,3\ell}) \ , 
\label{betaprime_3loop}
\eeq
where $a_{IR,3\ell}$ is the physical root of the quadratic equation above.  The
three-loop calculation in \cite{bc} used the value of $b_3$ in the
$\overline{\rm DR}$ scheme.  As mentioned above, 
we have found that the four-loop beta function does not exhibit a
physical IR zero over a substantial interval of $N_f$ in the NACP.  That is,
extracting the prefactor of $a^2$ in $\beta_{4\ell}$, we have found that the
cubic equation $b_1 + b_2 a + b_3 a^2+b_4a^3=0$ has no real positive zero in
this range of $N_f$.  We will discuss this further in the subsection on the LNN
limit.

In the special case $G={\rm SU}(N_c)$ and ${\cal R}=F$,
Eq. (\ref{betaprime_2loop}) reduces to
\beq
\beta'_{IR,2\ell} = \frac{N_c(3N_c-N_f)^2}{(2N_c^2-1)N_f-3N_c^2} \ . 
\label{betaprime_2loop_fund}
\eeq
To write an expression for the three-loop derivative, $\beta'_{IR,3\ell}$, it 
is convenient first to define two auxiliary polynomials in $N_c$ and $N_f$:
\beqs
D_s &=& -21N_c^5 + 21N_c^4N_f - 4N_c^3N_f^2-9N_c^2N_f \cr\cr
    &+& 3N_cN_f^2-2N_f
\label{ds}
\eeqs
and
\beqs
C_s &=& -54N_c^6+72N_c^5N_f-29N_c^4N_f^2+N_c^3N_f(4N_f^2-21) \cr\cr
&+&14N_c^2N_f^2-3N_cN_f(N_f^2+2)+3N_f^2 \ . 
\label{cs}
\eeqs
Then 
\beq
\beta'_{IR,3\ell} = \frac{N_c}{D_s}\Big [3N_c^3-2N_c^2N_f+N_f+\sqrt{C_s} \ 
\Big ] \ . 
\eeq
We will discuss these $n$-loop calculations further in the LNN limit
(\ref{lnn}) below. 


\subsection{Calculation via Series Expansion in $\Delta_f$} 

Proceeding the scheme-independent $\Delta_f$ expansion,
we calculate, for general $G$ and ${\cal R}$,
\beq
d_2 = \frac{2T_f^2}{3C_A C_f}
\label{d2}
\eeq
and
\beq
d_3 = \frac{2T_f^3(C_A+2C_f)}{(3C_AC_f)^2} \ . 
\label{d3}
\eeq
To our knowledge, these results are new. If $G={\rm SU}(N_c)$ and ${\cal
  R}=F$, then these take the form
\beq
{\rm SU}(N_c), \ R=F: \quad d_{2,F} = \frac{1}{3(N_c^2-1)}
\label{d2_fund}
\eeq
and
\beq
d_{3,F} = \frac{2N_c^2-1}{9N_c(N_c^2-1)^2} \ . 
\label{d3_fund}
\eeq
For this case of $G={\rm SU}(N_c)$ and ${\cal R}=F$, the next-higher-order 
coefficient is 
\beq
d_{4,F} = 
\frac{(N_c^4-2N_c^2+5)-18N_c^2(N_c^2+1)\zeta_3}{108N_c^2(N_c^2-1)^3} \ ,
\label{d4_fund}
\eeq
where $\zeta_s = \sum_{n=1}^\infty n^{-s}$ is the Riemann zeta function. 
These results for $d_j$, $2 \le j \le 4$ for $G={\rm SU}(N_c)$ and ${\cal R}=F$
agree with equivalent quantities given in \cite{gg}.  From these results for
$d_j$, $2 \le j \le 4$, it is evident that the coefficients $d_j$ in 
expansion (\ref{betaprime_ir_Deltaseries}) for $\beta'_{IR}$ does not have the
form of a geometric series.  This is in contrast to our theorem above and the
resultant Eq. (\ref{fjgeneral}) for the coefficient 
$f_{\Phi_{\rm prod},j}$ in expansion of the anomalous dimension of a
composite chiral superfield $\Phi_{\rm prod}$ in powers of $\Delta_f$, which
showed that the latter series is a geometric series. This is completely
consistent with our theorem, since the Konishi
supercurrent is not a (composite) chiral superfield.

The coefficients $d_2$ and $d_3$ are manifestly positive for any $G$ and $R$.
We find that $d_4$ is negative for all physical $N_c \ge 2$.  These are
qualitatively the same results that we found in \cite{dex} for
non-supersymmetric theories, namely that for arbitrary $G$ and $R$, $d_2$ and
$d_3$ are positive and in the case $G={\rm SU}(N_c)$ and ${\cal R}=F$, $d_4$ is
negative for all $N_c \ge 2$.  

The perfect agreement that we have found between the $\kappa_j$ that we have
calculated and the exact result (\ref{gamma_ir_sgt}) suggests that the same
agreement could hold for the $d_j$ with $1 \le j \le 3$ that we have
calculated. That is, these should also agree with the $d_j$ coefficients
obtained from the expansion of the exact $\beta'_{IR}$ as a series in powers of
$\Delta_f$ as expressed in Eq. (\ref{betaprime_ir_Deltaseries}). The only
difference is that in contrast to $\gamma_{M}$, one does not have an exact
closed-form expression for $\beta'_{IR}$ with which to compare in this ${\cal
  N}=1$ supersymmetric gauge theory.

In Table \ref{betaprime_values} we list the (scheme-independent) values that we
calculate for $\beta'_{IR,F,\Delta_f^p}$ with $2 \le p \le 4$ for the
illustrative gauge groups $G={\rm SU}(2)$, SU(3), and SU(4), as functions of
$N_f$ in the respective non-Abelian Coulomb phase intervals given in
Eq. (\ref{nacp}).
Numerically, 
\begin{widetext}
\beqs
{\rm SU}(2): \quad \beta'_{IR,F,\Delta_f^4} & =& \Delta_f^2 \Big [
  0.11111 + (4.3210 \times 10^{-2})\Delta_f
- (3.5986 \times 10^{-2})\Delta_f^2 \ \Big ] \cr\cr
& &
\label{betaprime_sunf_p4_su2}
\eeqs
\beqs
{\rm SU}(3): \quad \beta'_{IR,F,\Delta_f^4} & =& \Delta_f^2 \Big [
  4.1667 \times 10^{-2} + (0.98380 \times 10^{-2})\Delta_f
- (3.7763 \times 10^{-3})\Delta_f^2 \ \Big ] \cr\cr
& &
\label{betaprime_sunf_p4_su3}
\eeqs
and
\beqs
{\rm SU}(4): \quad \beta'_{IR,F,\Delta_f^4} & =& \Delta_f^2 \Big [
  2.2222 \times 10^{-2} + (3.8272 \times 10^{-3})\Delta_f
- (0.96987 \times 10^{-3})\Delta_f^2 \ \Big ] \ . \cr\cr
& &
\label{betaprime_sunf_p4_su4}
\eeqs
\end{widetext}
where the numerical coefficients are listed to the given floating-point
accuracy.

In Figs. \ref{betaprime_Nc2_susy_fig}-\ref{betaprime_Nc4_susy_fig} we show
plots of $\beta'_{IR,F,\Delta_f^p}$ with $2 \le p \le 4$ for these three
theories for $N_f$ in the respective non-Abelian Coulomb phase interval,
$(3/2)N_c < N_f < 3N_c$. (The plots also show the behavior for $N_f$ values 
slightly below the lower end of the NACP.) 

We next address the question of how well, for a given $G$, ${\cal R}$, and
$N_f$, the $\Delta_f$ expansion for $\beta'_{IR}$ converges in this ${\cal
  N}=1$ supersymmetric gauge theory.  We had carried out a similar analysis for
the $\Delta_f$ expansions for $\gamma_{M}$ and $\beta'_{IR}$ in our previous
work \cite{gtr}-\cite{dexl}. The $\Delta_f$ expansion is a series expansion
about $\Delta_f =0$, i.e., $N_f=N_u$, at the upper end of the non-Abelian
Coulomb phase. As $\Delta_f$ increases, i.e., as $N_f$ decreases below $N_u$,
one needs progressively more terms in this expansion to obtain an accurate
estimate of a given quantity.  In general, if $f(z)$ is an analytic function at
$z=0$, then it has a Taylor series expansion
\beq
f(z) = \sum_{j=1}^\infty f_j z^j \ .
\label{fz}
\eeq
The radius of convergence of this series, $z_c$, can be determined by the 
ratio test as 
\beq
z_c = \lim_{j \to \infty} \frac{|f_{j-1}|}{|f_j|} \ .
\label{zc}
\eeq
With the $\Delta_f$ expansion for $\beta'_{IR}$ 
considered as a Taylor series expansion, one could, in principle, calculate 
the radius of convergence ($cv$), $|\Delta_{f{,\rm cnv}}|$ as 
\beq
\Delta_{f,{\rm cnv}} = \lim_{j \to \infty} \frac{|d_j|}{|d_{j+1}|} \ . 
\label{radius}
\eeq
Clearly, it is not possible to apply this test precisely here for $\beta'_{IR}$
as a series in powers of $\Delta_f$, since one does not know the $d_j$ for $j
\to \infty$. Nevertheless, one can obtain a rough estimate of the radius of
convergence by calculating the ratios of adjacent coefficients for the first
few $d_j$.  We define the estimate of the radius of convergence given by the
ratio $|d_j/d_{j+1}|$ as
\beq
\Delta_{f,{\rm cnv},(j,j+1)} = \frac{|d_j|}{|d_{j+1}|} \ . 
\label{deltaconvjjplus1}
\eeq
Correspondingly, for a given $G$ and $R$, the minimum value of $N_f$ to which
the small-$\Delta_f$ expansion would be estimated to be convergent (denoted $mc$
for ``minimum ($N_f$) for convergence'') is
\beq
N_{f,{\rm mc},(j,j+1)} = N_u - \Delta_{f,{\rm cnv},(j,j+1)} \ , 
\label{nfminconv}
\eeq
where $N_u$ was given in Eq. (\ref{Nfb1z}). For general $G$ and ${\cal R}$, we
have
\beq
\Delta_{f,{\rm cnv},(j,j+1)} = \frac{d_2}{d_3} = 
\frac{3C_AC_f}{T_f(C_A+2C_f)} \ , 
\label{ratio23general}
\eeq
and hence
\beq
N_{f,{\rm mc},(2,3)} = \frac{3C_A^2}{2T_f(C_A+2C_f)} \ . 
\label{nfminconv23}
\eeq
This may lie above or below the lower end of the non-Abelian Coulomb phase at
$N_\ell$, as determined by the difference
\beq
N_{f,mc,(2,3)}-N_\ell = \frac{3C_A(C_A-2C_f)}{4T_f(C_A+2C_f)} \ . 
\label{nfminconv23_minus_nell}
\eeq
For example, for $G={\rm SU}(N_c)$, this difference is positive for the 
fundamental representation, but negative for the adjoint representation. 

We now focus on the case of main interest here, namely $G={\rm SU}(N_c)$ and 
${\cal R}=F$. For this case, 
\beq
\frac{d_{2,F}}{d_{3,F}} = \frac{3N_c(N_c^2-1)}{2N_c^2-1} \ , 
\label{ratio23_fund}
\eeq
so that 
\beq
N_{f,{\rm mc},(2,3)} = \frac{3N_c^3}{2N_c^2-1} \ . 
\label{nfminconv23_fund}
\eeq
Parenthetically, we observe that this difference is equal to the special case
of $N_{f,b2z}$ (given in general in Eq. (\ref{nfb2z})) for $G={\rm SU}(N_c)$
and ${\cal R}=F$. The value $N_{f,mc,(2,3)}$ lies above the lower end of the
non-Abelian Coulomb phase, as is evident from the difference
\beq
N_{f,{\rm mc},(2,3)}-N_\ell = \frac{3N_c^2}{2(2N_c^2-1)} \ . 
\label{nfminconv23_minus_nell_fund}
\eeq
As $N_c \to \infty$, this difference approaches zero.  

For the ratio of the next higher-order coefficients, we find 
\beq
\frac{d_{3,F}}{|d_{4,F}|} = \frac{12N_c(N_c^2-1)(2N_c^2-1)}
{ 18N_c^2(N_c^2+1)\zeta_3 -(N_c^4-2N_c^2+5)} \ , 
\label{ratio34_fund}
\eeq
so 
\beq
N_{f,{\rm mc},(3,4)}=
\frac{3N_c [ 18N_c^2(N_c^2+1)\zeta_3 -9N_c^4 +14N_c^2 -9 ]}
{ 18N_c^2(N_c^2+1)\zeta_3 -(N_c^4-2N_c^2+5)} \ . 
\label{nfminconv34_fund}
\eeq
This value lies above the lower end of the NACP, as is evident from the
difference 
\beqs
&&N_{f,{\rm mc},(3,4)}-N_\ell = \cr\cr
&& \frac{3N_c [ 18N_c^2(N_c^2+1)\zeta_3 
-17N_c^4 +26N_c^2 -13 ]}{ 18N_c^2(N_c^2+1)\zeta_3 -(N_c^4-2N_c^2+5)} \ . 
\cr\cr
&&
\label{nfminconv34_minus_nell_fund}
\eeqs
In Table \ref{nfvalues} we list values of $N_\ell$, $N_u$, $N_{f,{\rm
    mc},(2,3)}$, $N_{f,{\rm mc},(2,3)}-N_\ell$, $N_{f,{\rm mc},(3,4)}$, and
$N_{f,{\rm mc},(3,4)}-N_\ell$ for the illustrative cases $N_c=2, \ 3, \ 4$.
Thus, our analysis of the first two ratios of coefficients in the
small-$\Delta_f$ series expansion for $\beta'_{IR}$ suggests that the
small-$\Delta_f$ expansion for $\beta'_{IR}$ may be reliable over a substantial
portion of the non-Abelian Coulomb phase, including, in particular, the upper
and middle parts. In general, one would not expect the small-$\Delta_f$
expansion to apply reliably for small values $N_f$, where the properties of the
theory are qualitatively different from the properties in the non-Abelian
Coulomb phase.

These results on the convergence of the small-$\Delta_f$ expansion
(\ref{betaprime_ir_Deltaseries}) for $\beta'_{IR}$ may be compared with our
results for the convergence of the corresponding expansion
(\ref{gamma_ir_Deltaseries}) for $\gamma_{M}$.  As recalled above, we found
from our calculation of the coefficients $\kappa_j$ in the latter expansion
that they agreed perfectly with the Taylor series expansion of the exact result
(\ref{gamma_ir_sgt}).  This Taylor series expansion of (\ref{gamma_ir_sgt})
converges throughout the entire non-Abelian Coulomb phase.  Superficially, from
the analysis of the coefficients $d_j$ with $j=2, \ 3, \ 4$ in the
small-$\Delta_f$ series expansion of $\beta'_{IR}$, one might infer that this
series expansion might not converge as rapidly as the small-$\Delta_f$
expansion for $\gamma_{M}$ \cite{gg}.  However, one would need more terms to
get a better estimate of the actual region of convergence of the series
expansion of $\beta'_{IR}$ in powers of $\Delta_f$.  Especially in view of our
proof above that the series expansion in powers of $\Delta_f$ of
the anomalous dimension $\gamma_{\Phi_{\rm prod}}$ converges throughout the
entirety of the non-Abelian Coulomb phase, we believe that it is plausible that
the same is true of the corresponding series for $\beta'_{IR}$.  

For general $G$ and ${\cal R}$, since $d_2$ and $d_3$ are positive,
$\beta'_{IR}$ increases (initially quadratically) from 0 as $\Delta_f$
increases from 0, i.e., as $N_f$ decreases below its upper bound from
asymptotic freedom, $N_u$.  In the class of theories with $G={\rm SU}(N_c)$ and
${\cal R}=F$, we have calculated the next higher-order coefficient, $d_{4,F}$
and have shown that it is negative for all physical $N_c$. It is of interest to
investigate the consequences of the fact that $d_{4,F}$ is negative, bearing in
mind the cautionary remarks concerning the range in $N_f$ in which the
small-$\Delta_f$ may be reasonably reliable.  Because $d_{4,F}$ is negative, as
$\Delta$ increases from 0, i.e., as $N_f$ decreases from $N_u$, the $d_4
\Delta_f^4$ term in $\beta'_{IR}$ eventually stops the initial increase in
$\beta'_{IR,\Delta_f^4}$ and, for smaller $N_f$, causes
$\beta'_{IR,\Delta_f^4}$ to decrease.  If one were to use the $\Delta_f$
expansion for sufficiently small values of $N_f$, then the series for
$\beta'_{IR}$ calculated to $O(\Delta_f^4)$, i.e., $\beta'_{IR,\Delta_f^4}$,
would pass through zero to negative values.  We first determine the value of
$\Delta_f$, or equivalently, $N_f$, at which $\beta'_{IR,\Delta_f^4}$
vanishes. The condition that $\beta'_{IR,\Delta_f^4}=0$ is the equation
\beq
\Delta_f^2( d_2 + d_3 \, \Delta_f + d_4\, \Delta_f^2)=0 \ . 
\label{betaprime_p4zero}
\eeq
Aside from the solution $\Delta_f=0$, i.e., $N_f=N_u$, this equation has two
solutions, corresponding to the roots of the quadratic factor.  Of these, we
denote the relevant one as $\Delta_0 = N_u - N_{f,0}$. We calculate 
\begin{widetext}
\beq
N_{f,0} = \frac{3N_c\Big [ N_c^4(-5+18\zeta_3) + 2N_c^2(4+9\zeta_3) - 7 
-2(N_c^2-1)\sqrt{S_0}}{N_c^4(18\zeta_3-1) + 2N_c^2(1+9\zeta_3) - 5} \ , 
\label{nfzero}
\eeq
\end{widetext}
where 
\beq
S_0 = 3N_c^4(1+6\zeta_3) + 2N_c^2(-1+9\zeta_3)-4 \ . 
\label{s0}
\eeq
(The other root of the quadratic factor, with the opposite sign in front of the
square root, is greater than $N_u$ and hence is not relevant here, since we
restrict $N_f < N_u$ for asymptotic freedom.)  Numerically, for the
illustrative values $N_c=2, \ 3, \ 4$, our expression for $N_{f,0}$ (understood
to be continued from the positive integers to the positive real numbers) takes
the respective values 3.5427, 4.1294, and 4.8496.  In these three cases, as is
evident from Table \ref{nfvalues}, $N_\ell$ has the respective values =3, \
4.5, \ 6, so that for ${\cal R}=F$ and $G={\rm SU}(2)$, $N_{f,0} > N_\ell$,
while for SU(3) and SU(4), $N_{f,0} < N_\ell$.

Using electric-magnetic duality, it has been concluded that for $G={\rm
  SU}(N_c)$ and ${\cal R}=F$, $\beta'_{IR}$ vanishes quadratically at the lower
end of the non-Abelian Coulomb phase at $N_f = N_\ell = (3/2)N_c$
\cite{grisaru97}:
\beq
\beta'_{IR} = \frac{28}{3} \, \Big ( \frac{N_f}{N_c} - \frac{3}{2} \Big )^2 
\quad {\rm as} \ N_f \searrow \frac{3N_c}{2} \ . 
\label{betaprime_Nell}
\eeq
Given the fact that our $\Delta_f$ expansion starts from the
other (i.e., the upper) end of the non-Abelian Coulomb phase, we would not
expect our calculations of $\beta'_{IR}$ to $O(\Delta_f^4)$ for this theory 
to precisely reproduce this behavior at $N_f=(3/2)N_c$.  
Taking this into account, our numerical results on $N_{f,0}$ are consistent 
with the behavior in (\ref{betaprime_Nell}). It should be noted that 
the three values listed above for $N_{f,0}$ actually lie below the minimum 
values where where our estimates indicate that the
small-$\Delta_f$ series is reliable, namely the values 4.8, 6.4, and 8.05 for
$N_c=2, \ 3, \ 4$, respectively, as listed in Table \ref{nfvalues}.  
A general statement is that our calculations of series
expansions for $\beta'_{IR}$ in both the nonsupersymmetric gauge theory
\cite{dex,dexs,dexl} and the results present here for the supersymmetric gauge
theory show qualitatively quite different behavior than we have found for both
$\gamma_{M}$ and $\gamma_{B}$.  In the latter two cases, all
of the coefficients in the small-$\Delta_f$ expansion are positive, leading to
the two monotonicity theorems mentioned above.


\section{Results in the Limit of Large $N_c$ and $N_f$ with $N_f/N_c$ Fixed }
\label{lnn_section}

\subsection{General} 

For this class of theories with $G={\rm SU}(N_c)$ and ${\cal R}=F$, an
interesting limit is
\beqs
& & LNN: \quad N_c \to \infty \ , \quad N_f \to \infty \cr\cr
& & {\rm with} \ r \equiv \frac{N_f}{N_c} \ {\rm fixed \ and \ finite}  \cr\cr
& & {\rm and} \ \ \xi(\mu) \equiv \alpha(\mu) N_c \ {\rm is \ a \
finite \ function \ of} \ \mu \ .
\cr\cr
& &
\label{lnn}
\eeqs
We will use the symbol $\lim_{LNN}$ for this limit, where ``LNN'' stands
for ``large $N_c$ and $N_f$'' with the constraints in Eq. (\ref{lnn})
imposed. This is sometimes called the 't Hooft-Veneziano limit.

We define the following quantities in this limit:
\beq
\xi = 4\pi x = \lim_{LNN} \alpha N_c \ ,
\label{xlnn}
\eeq
\beq
r_u = \lim_{LNN} \frac{N_u}{N_c} \ , 
\label{rb1zdef}
\eeq
and
\beq
r_\ell = \lim_{LNN} \frac{N_\ell}{N_c} \ , 
\label{rb2zdef}
\eeq
with values
\beq
r_u = 3 \ , \quad\quad  r_\ell = \frac{3}{2} \ . 
\label{rb12z}
\eeq
These quantities are listed in Table \ref{rvalues}. 
Thus, the non-Abelian Coulomb phase occurs for $r$ in the interval
\beq
{\rm LNN, \ \ NACP}: \quad \frac{3}{2} < r < 3 \ . 
\label{intervalr}
\eeq
We define the scaled scheme-independent expansion parameter for the LNN limit
\beq
\Delta_r \equiv \frac{\Delta_f}{N_c} = r_u-r = 3-r \ .
\label{deltar}
\eeq
As $r$ decreases from $r_u$ to $r_\ell$ in the non-Abelian Coulomb phase, 
$\Delta_r$ increases from 0 to a maximal value
\beqs
& & (\Delta_r)_{\rm max} = r_u-r_\ell = \frac{3}{2} 
\quad {\rm for} \ r \in {\rm NACP} \ . \cr\cr
& &
\label{Deltar_max_irz}
\eeqs
%


\subsection{$\gamma_{M}$ in the LNN Limit} 

For the analysis of $\gamma_{M}$ at the superconformal IRFP, we define rescaled
coefficients $\hat \kappa_{j,F}$
\beq
\hat \kappa_{j,F} \equiv \lim_{N_c \to \infty} N_c^j \, \kappa_{j,F}
\label{kappahatn}
\eeq
that are finite in this LNN limit. The anomalous dimension
$\gamma_{IR}$ is also finite in this limit and is given by
\beq
{\cal R}=F: \quad \lim_{LNN} \gamma_{M,LNN} = 
\sum_{j=1}^\infty \hat \kappa_{j,F} \Delta_r^j \ .
\label{gamma_ir_lnn}
\eeq
In this LNN limit, the exact result for $\gamma_{M,LNN}$ 
(\ref{gamma_ir_sgt}) takes the form 
\beq
\gamma_{M,LNN} = \frac{\frac{\Delta_r}{3}}
{1 - \frac{\Delta_r}{3}} \ , 
\label{gamma_ir_sgt_lnn}
\eeq
so that 
\beq
\hat\kappa_{j,F} = \frac{1}{3} \quad \forall \ j \ . 
\label{kappahatj}
\eeq
%


\subsection{Rescaled $\gamma_{B}$ in the LNN Limit} 

To construct a rescaled anomalous dimension at the superconformal IRFP that is
finite in the LNN limit, we define
\beq
\hat \gamma_{B} \equiv \lim_{LNN} \frac{\gamma_{B}}{N_c} \ ,
\label{gamma_baryon_rescaled_lnn}
\eeq
and similarly with $\hat\gamma_{\tilde B} = \hat\gamma_{B}$.  By
construction, these rescaled baryon anomalous dimensions are finite in the LNN
limit and have the common value
\beq
\lim_{LNN} \hat \gamma_{B} = \frac{1}{2}\Big ( \frac{3}{r}-1 \Big )  \ . 
\label{gamma_baryon_hat}
\eeq
%


\subsection{$\beta'_{IR}$ in the LNN Limit}

The rescaled beta function that is finite and nontrivial in the LNN limit is 
\beq
\beta_{\xi} \equiv \frac{d\xi}{dt} = \lim_{LNN} \beta_\alpha N_c \ ,
\label{betaxi}
\eeq
with the series expansion
\beq
\beta_\xi \equiv \frac{d\xi}{dt}
= -8\pi x \sum_{\ell=1}^\infty \hat b_\ell x^\ell
= -2 \xi \sum_{\ell=1}^\infty \tilde b_\ell \xi^\ell \ ,
\label{betaxiseries}
\eeq
where
\beq
  \hat b_\ell = \lim_{LNN} \frac{b_\ell}{N_c^\ell} \ , \quad
\tilde b_\ell = \lim_{LNN} \frac{\bar b_\ell}{N_c^\ell} \ .
\label{bellrel}
\eeq

The first two 
rescaled coefficients of the beta function, which are scheme-independent, are 
\beq
\hat b_1 = 3-r
\label{b1hat}
\eeq
and
\beq
\hat b_2 = 2(3-2r) \ . 
\label{b2hat}
\eeq
In the $\overline{\rm DR}$ scheme, 
\beq
\hat b_3 = 21-21r+4r^2
\label{b3hat}
\eeq
and 
\beq
\hat b_4 = 2[51-66r+3(7+2\zeta_3)r^2] \ . 
\label{b4hat}
\eeq
In the LNN limit, one has the scheme-independent two-loop result
\beq
x_{IR,2\ell} = -\frac{\hat b_1}{\hat b_2} = \frac{3-r}{2(2r-3)} \ . 
\label{xir_2loop}
\eeq

At the three-loop level, $x_{IR,3\ell}$ is the physical root among the two
roots of the quadratic equation $\hat b_1 + \hat b_2 x + \hat b_3 x^2=0$. It is
convenient to define two auxiliary polynomials:
\beq
C_r = \lim_{LNN} \frac{C_s}{N_c^6} = -54+72r-29r^2+4r^3
\label{cs_susy}
\eeq
and
\beq
D_r = \lim_{LNN} \frac{D_s}{N_c^5} = -21 + 21r -4r^2 \ .
\label{ds_susy}
\eeq
Then
\beq
x_{IR,3\ell} = \frac{2[ -(2r-3) + \sqrt{C_s} \ ]}{D_s} \ . 
\label{xir_3loop_susy}
\eeq

These inputs were used to calculate $\beta'_{IR}$ in the LNN limit
\cite{bc,lnn}.  At two-loop order, one has the scheme-independent result,
\beq
\beta'_{\xi,IR,2\ell} = \frac{(3-r)^2}{2r-3} \ . 
\label{betaprime_2loop_lnn}
\eeq

At the three-loop order \cite{bc,lnn} 
\beqs
&&\beta'_{\xi,IR,3\ell} = 2x_{IR,3\ell}(2b_1 + b_2 x_{IR,3\ell}) \cr\cr
&=& 2(3-r)\Big [ -(2r-3)+\sqrt{C_r} \ \Big ] \times \cr\cr
&\times&\bigg [ 1 + \frac{2\Big [-(2r-3)+\sqrt{C_r} \ \Big ]}{D_s} \ \bigg ] 
\ . 
\label{betaprime_3loop_lnn}
\eeqs
We list the values of $\beta'_{IR,2\ell}$ and $\beta'_{IR,3\ell}$ in Table
\ref{betaprime_values_lnn}. 

We find that the four-loop beta function does not exhibit a physical (i.e.,
real, positive) IR zero over a substantial portion of the NACP interval $3/2 <
r < 3$.  Specifically, extracting the prefactor proportional to $x^2$ in
$\beta_{\xi,4\ell}$, we find that, as $r$ decreases from its upper bound of
$r=3$ in the NACP, the equation $\hat b_1 + \hat b_2 x + \hat b_3 x^2 + \hat
b_4=0$ ceases to exhibit a physical zero as $r$ decreases below $r_0 = 2.6165$.
We recall that we found that although the $n$-loop beta function had a physical
IR zero for $n=2$, 3, and 4 loops in the corresponding nonsupersymmetric
SU($N_c$) theory with $N_f$ fermions with ${\cal R}=F$, this was not the case
at the five-loop level \cite{flir}, and in the LNN limit, as $r$ decreased
below its upper limit of 5.5, the five-loop beta beta function ceased to
exhibit a physical IR zero as $r$ decreased through the value $r_{cx}=4.3226$
(as given in Eq. (5.3) of \cite{dexl}).  Thus, this complication appears at one
loop lower (i.e. at the four-loop level) in the present supersymmetric theory,
as compared with the case of the nonsupersymmetric theory with the same $G$ and
$R$.  This shows again the advantage of the scheme-independent expansion
method, since it does circumvents the explicit extraction of
$\alpha_{IR,n\ell}$ (here, $x_{n,\ell}$ in the LNN limit) in order to calculate
values of physical quantities at the IRFP.

For the scheme-independent expansion, in addition to the 
rescaled quantity $\Delta_r$ defined in Eq. (\ref{deltar}), we define the
rescaled coefficient
\beq
\hat d_{j,F} = \lim_{LNN} N_c^j \, d_{j,F} \ ,
\label{dnhat}
\eeq
which is finite.  Then each term
\beqs
\lim_{LNN} d_{j,F} \Delta_r^j &=& \lim_{LNN} (N_c^j d_{j,F})
\Big ( \frac{\Delta_r}{N_c} \Big )^j \cr\cr
& = & \hat d_{j,F} \Delta_r^j \ . 
\label{finiteproduct}
\eeqs
is finite in this limit.
Thus, writing $\lim_{LNN} \beta'_{IR}$ as $\beta'_{IR,LNN}$ for this
${\cal R}=F$ case, we have
\beq
\beta'_{IR,LNN} = \sum_{j=1}^\infty \hat d_{j,F} \Delta_r^j \ . 
\label{betaprime_ir_lnn}
\eeq

From our results (\ref{d2_fund}), (\ref{d3_fund}), and (\ref{d4_fund}), it
follows that
\beq
\hat d_{2,F} = \frac{1}{3} \ , 
\label{d2hat_lnn}
\eeq
\beq
\hat d_{3,F} = \frac{2}{9} = 0.22222 \ , 
\label{d3hat_lnn}
\eeq
and
\beq
\hat d_{4,F} = -\frac{(18\zeta_3-1)}{108} = -0.19108 \ . 
\label{d4hat_lnn}
\eeq
Thus, in this LNN limit, to $O(\Delta_r^4)$, 
\beq
\beta'_{IR,LNN,\Delta_r^4} = \Delta_r^2 \Big [  \frac{1}{3} + 
\frac{2}{9} \, \Delta_r - \Big ( \frac{18\zeta_3-1}{108}\Big ) \, \Delta_r^2 
\Big ] \ . 
\label{betaprime_lnn_p4}
\eeq

In Table \ref{betaprime_values_lnn} we list the (scheme-independent) values
that we calculate for $\beta'_{IR,LNN,\Delta_r^p}$
and in Fig. \ref{betaprime_lnn_susy_fig}, we plot $\beta'_{IR,LNN,\Delta_r^p}$
with $2 \le p \le 4$, as functions of $r$ in the non-Abelian Coulomb phase 
interval $3/2 < r < 3$. (The plot also shows the behavior slightly below the
lower end of the NACP.) 

To obtain a rough estimate of the interval in $r$ in which this
small-$\Delta_r$ expansion is reliable, we follow the same procedure as before
for finite $N_c$ and $N_f$.  Analogously to Eqs. (\ref{deltaconvjjplus1}) and
(\ref{nfminconv}), we define
\beq
(\Delta_r)_{{\rm cnv},(j,j+1)} = \frac{|\hat d_j|}{|\hat d_{j+1}|} 
\label{deltarconvjjplus1}
\eeq
and 
\beq
r_{{\rm mc},(j,j+1)} = r_u - \Delta_{r,{\rm cnv},(j,j+1)} \ . 
\label{rminconv}
\eeq
We calculate 
\beq
(\Delta_r)_{{\rm cnv},(2,3)} = \frac{3}{2} 
\label{d23ratio_lnn}
\eeq
and 
\beq
(\Delta_r)_{cnv,(3,4)} = \frac{24}{18\zeta_3-1} = 1.16296 \ , 
\label{d34ratio_lnn}
\eeq
so that
\beq
r_{{\rm mc},(2,3)} = \frac{3}{2}
\label{rminconv23}
\eeq
and
\beq
r_{{\rm mc},(3,4)} = \frac{27(2\zeta_3-1)}{18\zeta_3-1} = 1.8370 
\label{rminconv34}
\eeq
Since the lower end of the non-Abelian Coulomb phase occurs at $r_\ell=3/2$,
this analysis suggests that the small-$\Delta_r$
expansion may be reasonably reliable for a substantial part of this phase, 
extending down from $r=3$ to around $r \sim 1.8$.

In the present LNN limit, the condition that $\beta'_{IR,LNN}=0$ is satisfied
at $\Delta_r=0$, i.e., $r=3$, and at the relevant solution of the quadratic
factor in Eq. (\ref{betaprime_lnn_p4}).  We define
\beq
\Delta_{r,0} = 3-r_0 \ , 
\label{Delta_r0}
\eeq
with
\beq
r_0 = \lim_{LNN} \frac{N_{f,0}}{N_c} \ . 
\label{r0}
\eeq
We calculate 
\beq
r_0 = \frac{3\Big [18\zeta_3-5-2\sqrt{3(1+6\zeta_3)} \ \Big ] }
{18\zeta_3-1} = 0.975415 \ , 
\label{r0_lnn}
\eeq
where the numerical value is given to the indicated floating-point accuracy.
(The other root of the quadratic, with the opposite sign in front of the square
root, is $r=3.861627$, which is greater than $r_u=3$ and hence is not
relevant.)  Evidently, $r_0$ is less $r_\ell=1.5$, i.e., it lies below the
lower end of the non-Abelian Coulomb phase and well below the region in $r$
where the small-$\Delta_r$ expansion is expected to be reliable, based on our
analysis of ratios of $\hat d_j$ above.  

In the LNN limit, the result (\ref{betaprime_Nell}) from \cite{grisaru97} is 
\beq
\beta'_{IR} = \frac{28}{3}\Big ( r - \frac{3}{2} \Big )^2 \quad {\rm as} \ 
r \searrow \frac{3}{2} \ .
\label{betaprime_rell}
\eeq
Analogously to our discussion above for finite $N_c$ and $N_f$, here in the LNN
limit, given that the $\Delta_r$ series expansion for $\beta'_{IR}$ starts from
the other end of the NACP at $r=3$, i.e., $\Delta_r=0$, we would not anticipate
that our series expansion to $O(\Delta_r^4)$ would closely reproduce this
property of $\beta'_{IR}$.  With the calculation of $\beta'_{IR}$ to
$O(\Delta_r^4)$, we may observe at least that as $r$ decreases toward the lower
end of the NACP, $\beta'_{IR}$ curves over and decreases, approaching the zero
at $r_0$.  As was true of our analysis for finite $N_c$ and $N_f$, given the
limited order in the $\Delta_r$ series expansion to which we have calculated
$\beta'_{IR}$ and our estimate of the region over which this expansion may be
used reliably, we consider that our results are consistent with the behavior
(\ref{betaprime_rell}). 

In view of (\ref{betaprime_rell}), it is of interest to study a structural form
for $\beta'_{IR,LNN}$ that incorporates a double zero at $r=3/2$, via the
factor $[1-(2/3)\Delta_r]^2$ as well as the double zero at $r=3$, as embodied
in the factor $(3-r)^2 = \Delta_r^2$.  We thus write
\beq
\beta'_{IR} = \Delta_r^2 \, [1-(2/3)\Delta_r]^2 \, \Big [ \hat h_2 +
\hat h_3 \Delta_r + \hat h_4 \Delta_r^2 + O(\Delta_r^3) \ \Big ] 
\label{betaprime_dualform}
\eeq
The coefficients $\hat h_j$ are related to the $\hat d_j$ that we have
calculated as follows: 
\beq
\hat h_2 = \hat d_2 = \frac{1}{3} 
\label{h2hat}
\eeq
\beq
\hat h_3 = \hat d_3 + \frac{4}{3}\hat h_2 = \frac{2}{3}
\label{h3hat}
\eeq
\beqs
\hat h_4 &=& \hat d_4 + \frac{4}{3}\hat h_3 - \frac{4}{9}\hat h_2 \cr\cr
         &=& \frac{9-2\zeta_3}{12} \ . 
\label{h4hat}
\eeqs
Calculations to higher order in $\Delta_r$ would be necessary in order to
reproduce the coefficient (28/3) in Eq. (\ref{betaprime_rell}).


\subsection{Pad\'e Approximants for $\beta'_{IR}$ in the LNN Limit}

It is also of interest to calculate and analyze Pad\'e approximants. For this
purpose, it is convenient to define a reduced (red.) function normalized to be
equal to unity at $\Delta_r=0$: 
\beq
\beta'_{IR,LNN,red.} = \frac{\beta'_{IR,LNN}}{\hat d_2 \Delta_r^2} = 
 1 + \frac{1}{\hat d_2} \sum_{j=3}^\infty \hat d_j \Delta_r^{j-2} \ . 
\label{betaprime_lnn_reduced}
\eeq
Thus, from $\beta'_{IR,LNN,\Delta_r^5}$, we have 
\beqs
\beta'_{IR,LNN,red.} &=& 1 + \frac{2}{3}\Delta_r - \frac{(18\zeta_3-1)}{36} \,
\Delta_r^2 + O(\Delta_r^3) \cr\cr
&=& 1 + 0.66667 \Delta_r - 0.57325\Delta_r^2 + O(\Delta_r^3) \ . \cr\cr
& & 
\label{betaprime_lnn_reduced_p4}
\eeqs
We recall that the $[p,q]$ Pad\'e approximant to a finite Taylor series $f(x)=1+\sum_{n=1}^m
x^n$ is the rational function
\beq
f_{[p,q]} = \frac{1+\sum_{j=1}^p \, n_j x^j}{1+\sum_{k=1}^q d_k \, x^k}
\label{pqx}
\eeq
with $p+q=m$, where the coefficients $n_j$ and $d_k$ are independent of
$x$. Thus, in the present case, with $x=\Delta_r$ and
$f(\Delta_r)=\beta'_{IR,LNN,red.}$, calculated to $O(\Delta_r^2)$
(corresponding to the calculation of $\beta'_{IR,LNN}$ to $O(\Delta_r^4)$), it
follows that, aside from the Pad\'e approximant [2,0], which is the function
$\beta'_{IR,LNN,red.}$, itself, there are two Pad\'e approximants to the
series, namely [1,1] and [0,2].  For the first of these, we calculate
\beq
\beta'_{IR,LNN,red.,[1,1]} = \frac{1+ \frac{1}{8}(5+6\zeta_3)\Delta_r}
{1+\frac{1}{24}(18\zeta_3-1)\Delta_r} \ . 
\label{betaprime_reduced_lnn_p4_pade11}
\eeq
The pole in this [1,1] Pad\'e approximant occurs at
\beq
(\Delta_r)_{[1,1],{\rm  pole}} = -\frac{24}{18\zeta_3-1} = -1.162958 \ , 
\label{betaprime_reduced_lnn_p4_pade11_rpole}
\eeq
where the numerical value is given to the indicated floating-point accuracy.
Hence, this Pad\'e approximant converges in a disk centered at $\Delta_r=0$ in
the complex $\Delta_r$ plane of radius 1.162958.  This does not cover all of
the non-Abelian Coulomb phase, but does extend down to $r=1.8370$, close to the
lower boundary of the NACP at $r=1.5$.  This [1,1] Pad\'e approximant does not
have any zero in the NACP; its zero occurs at $\Delta_r=-8/(5+6\zeta_3)$, or
equivalently, in terms of $r$, at 
\beq
r_{[1,1],{\rm zero}} = \frac{23+18\zeta_3}{5+6\zeta_3} = 3.6551 \ . 
\label{betaprime_reduced_lnn_p4_pade11_rzero}
\eeq
Evidently, this zero lies above the upper end of the NACP at $r_u=3$ 
(but within the radius of convergence of the approximant).  

For the [0,2] Pad\'e approximant to $\beta'_{IR,LNN,red.}$, we calculate
\beq
\beta'_{IR,LNN,red.,[0,2]} = \frac{1}
{1-\frac{2}{3}\Delta_r + \frac{1}{12}(5+6\zeta_3)\Delta_r^2} \ . 
\label{betaprime_reduced_lnn_p4_pade02}
\eeq
The poles of the approximant occur at the complex-conjugate points 
\beq
(\Delta_r)_{[0,2],{\rm pole}} = \frac{2(2 \pm i\sqrt{11+18\zeta_3} \ )} 
{5+6\zeta_3} \ . 
\label{betaprime_reduced_lnn_p4_pade02_poles}
\eeq
These have magnitude 
\beq
|(\Delta_r)_{[0,2],{\rm pole}}| = \frac{2\sqrt{3}}{\sqrt{5+6\zeta_3}} = 
0.991268 \ , 
\label{betaprime_reduced_lnn_p4_pade02_poles_mag}
\eeq
so that this [0,2] Pad\'e approximant converges for $\Delta_r$ in the disk
defined by $|\Delta_r| < 0.991268$ in the complex $\Delta_r$ plane.  
On the real axis, this disk
of convergence extends down to $r=2.0087$ and hence covers about 2/3 of the
non-Abelian Coulomb phase interval $3/2 < r < 3$. 

Although a $[p,q]$ Pad\'e approximant only contains information about a
function up to the highest-order term that has been calculated, namely the
$O(\Delta_r^2)$ term in $\beta'_{IR,LNN,red.}$ (equivalently, the 
$O(\Delta_r^4)$ term in $\beta'_{IR,LNN}$), it is of interest to investigate
the series expansion of such an approximant with $q \ne 0$, going to higher
order. This can sometimes give a hint about the next-higher order term in the
Taylor series expansion for the original function.  In the present case, we
calculate the expansions
\beqs
&& \beta'_{IR,LNN,red.,[1,1]} = 1 + \frac{2}{3}\Delta_r - 
\Big (\frac{18\zeta_3-1}{2^2 \cdot 3^2} \Big ) \, \Delta_r^2 \cr\cr
&&+ \frac{(18\zeta_3-1)^2}{2^5 \cdot 3^3} \, \Delta_r^3 + O(\Delta_r^4) \cr\cr
&&= 1 + 0.66667\Delta_r -0.57325\Delta_r^2 + 0.49292\Delta_r^3 + O(\Delta_r^4) 
\cr\cr
&& 
\label{betaprime_pade11_taylor}
\eeqs
and
\beqs
&& \beta'_{IR,LNN,red.,[0,2]} = 1 + \frac{2}{3}\Delta_r - 
\Big ( \frac{18\zeta_3-1}{2^2 \cdot 3^2}\Big ) \, \Delta_r^2 \cr\cr
&&-\Big ( \frac{18\zeta_3+7}{3^3} \Big ) \, \Delta_r^3 +O(\Delta_r^3) \cr\cr
&&= 1 + 0.66667\Delta_r -0.57325\Delta_r^2 - 1.06063\Delta_r^3 + O(\Delta_r^4)
\ . 
\cr\cr
&&
\label{betaprime_pade02_taylor}
\eeqs
Since the sign of the $O(\Delta_r^3)$ term of $\beta'_{IR,LNN,red.}$
(equivalent to the sign of $d_5$, since ${\rm sgn}(d_1) > 0$) predicted by the
Taylor series expansion of $\beta'_{IR,LNN,red.,[1,1]}$ is positive, which is
opposite to the negative-sign prediction of the Taylor series expansion of
$\beta'_{IR,LNN,red.,[0,2]}$, these expansions do not give any consistent hint
of the sign of $\hat d_5$.

In this context, one may ask what the analogous calculations would have yielded
in the case of a nonsupersymmetric SU($N_c$) gauge theory in the same LNN
limit.  In our previous analyses \cite{dexs,dexl} we had already gone beyond
this stage and calculated the actual $d_5$ coefficient and thus 
$\beta'_{IR}$ to $O(\Delta_f^5)$.
However, since we do not have $b_5$ available in the supersymmetric theory, in
contrast to the nonsupersymmetric theory, there is a motivation here to see
what the Taylor series expansions of the Pad\'e approximants to $\beta'_{IR}$,
calculated to $O(\Delta_r^4)$ would have suggested about the possible sign of
the next-higher-order coefficient, $\hat d_5$.  Thus, we calculate Pad\'e
approximants to the reduced function defined in Eq. 
(\ref{betaprime_lnn_reduced}) defined to be unit-normalized at $\Delta_r=0$.
From our results in \cite{dex,dexs,dexl} we have
\begin{widetext}
\beqs
\beta'_{IR,LNN,red.,ns.} &=& 1 + \frac{26}{3 \cdot 5^2}\Delta_r + 
\Big ( \frac{366782}{3^3 \cdot 5^8} -\frac{352}{3^2 \cdot 5^4}\zeta_3 
\Big )\Delta_r^2 \cr\cr
&+& \Big ( -\frac{596389102}{3^4 \cdot 5^{12}} - 
\frac{90304}{3^3 \cdot 5^7}\zeta_3 +\frac{22528}{3^3 \cdot 5^6}\zeta_5 \Big ) 
\Delta_r^3 + O(\Delta_r^4) \cr\cr
&=& 1 + 0.34667\Delta_r - 0.040446\Delta_r^2 - 0.0262475\Delta_r^3 + 
O(\Delta_r^4) \ , 
\label{betaprime_lnn_reduced_nonsusy}
\eeqs
\end{widetext}
where the subscript $ns.$ stands for ``nonsupersymmetric''. 
Our format here and below is to indicate the simple factorizations of the
denominators of the various terms. In general, the numerators do not have such
simple factorizations; for example, $366782=2 \cdot 13 \cdot 14107$, etc.
Now we calculate the [1,1] and [0,2] Pad\'e approximants to the truncation of 
$\beta'_{IR,LNN,red.,ns.}$ to $O(\Delta_r^2)$.  These are
\beqs
&& \beta'_{IR,LNN,red.,ns.,[1,1]}=\frac{1+\Big (\frac{34643}{3^2 \cdot 5^6} +
\frac{176}{3 \cdot 5^2 \cdot 13}\zeta_3 \Big )\Delta_r}
{1 + \Big ( -\frac{14107}{3^2 \cdot 5^6} + 
\frac{176}{3 \cdot 5^2 \cdot 13}\zeta_3 \Big )\Delta_r } 
\cr\cr
&&
\label{betaprime_lnn_reduced_nonsusy_pade11}
\eeqs
and
\beqs
&& \beta'_{IR,LNN,red.,ns.,[0,2]} = \cr\cr
&& \frac{1}{1 -\frac{26}{3 \cdot 5^2}\Delta_r 
+ \Big (\frac{900718}{3^3 \cdot 5^8}+
\frac{352}{3^2 \cdot 5^4}\zeta_3 \Big )\Delta_r^2 } \ .
\label{betaprime_lnn_reduced_nonsusy_pade02}
\eeqs
Next, we expand these in Taylor series around $\Delta_r=0$ to see what they
predict for the $O(\Delta_r^3)$ term $(\hat d_5/{\hat d_2})\Delta_r^3$ in
$\beta'_{IR,LNN,red.,ns.}$, or equivalently, the $O(\Delta_r^5)$ term in 
$\beta'_{IR,LNN,ns.}$.  We thus ascertain how these predictions compare with
the actual $O(\Delta_r^5)$ term that we have calculated in 
$\beta'_{IR,LNN,ns.}$ in \cite{dexs,dexl}.  We have 
\begin{widetext}
\beqs
\beta'_{IR,LNN,red.,ns.,[1,1]} &=& 1 + \frac{26}{3 \cdot 5^2}\Delta_r + 
\Big ( \frac{366782}{3^3 \cdot 5^8} -\frac{352}{3^2 \cdot 5^4}\zeta_3 
\Big )\Delta_r^2 
+\Big ( \frac{366782}{3^3 \cdot 5^8} - \frac{352}{3^2 \cdot 5^4}\zeta_3 
\Big ) \Big ( \frac{14107}{3^2 \cdot 5^6} - 
\frac{176}{3 \cdot 5^2 \cdot 13}\zeta_3 \Big )\Delta_r^3 +O(\Delta_r^4) \cr\cr
&=&1 + 0.34667\Delta_r - 0.040446\Delta_r^2 + 0.0047188\Delta_r^3 + 
O(\Delta_r^4)
\label{betaprime_lnn_reduced_nonsusy_pade11_taylor}
\eeqs
and
\beqs
\beta'_{IR,LNN,red.,ns.,[0,2]} &=& 1 + \frac{26}{3 \cdot 5^2}\Delta_r + 
\Big ( \frac{366782}{3^3 \cdot 5^8} -\frac{352}{3^2 \cdot 5^4}\zeta_3 
\Big )\Delta_r^2 
-\Big ( \frac{13882336}{3^4 \cdot 5^{10}} + 
\frac{18304}{3^3 \cdot 5^6}\zeta_3 \Big ) 
\Delta_r^3 + O(\Delta_r^4) \cr\cr
&=&1 + 0.34667\Delta_r - 0.040446\Delta_r^2 -0.0697040\Delta_r^3 + 
O(\Delta_r^4) \ . 
\label{betaprime_lnn_reduced_nonsusy_pade02_taylor}
\eeqs
\end{widetext}
The terms up to $O(\Delta_r^2)$, must, of course, coincide with the
corresponding terms in $\beta'_{IR,LNN,red.,ns.}$.  We find that the Taylor
series expansions of $\beta'_{IR,LNN,red.,ns.,[1,1]}$ and
$\beta'_{IR,LNN,red.,ns.,[0,2]}$ yield respective $O(\Delta_r^3)$ terms with
signs that are opposite to, and the same as, the actual $O(\Delta_r^3)$ term in
$\beta'_{IR,LNN,red.,ns.}$ that we calculated in \cite{dexs,dexl}, shown above
in Eq. (\ref{betaprime_lnn_reduced_nonsusy}).  Hence, if this nonsupersymmetric
case is a guide to the situation in the supersymmetric theory considered here,
then our Taylor series expansion of the $\beta'_{IR,LNN,red.,[0,2]}$ in the
supersymmetric theory (Eq. (\ref{betaprime_pade02_taylor}) may be
expected to yield the correct sign of the $O(\Delta_r^3)$ term in
$\beta'_{IR,LNN,red.}$, or equivalently, the $O(\Delta_r^5)$ term in
$\beta'_{IR,LNN}$, i.e., the sign of $\hat d_5$.  Thus, this predicts that the
sign of $\hat d_5$ is negative.  We emphasize, however, that this
procedure is obviously nonrigorous, since these Pad\'e approximants in the
supersymmetric theory only contain information from the 
$\hat d_j$ with $j=2, \ 3, \ 4$.


\section{Conclusions}
\label{conclusion_section}

In this paper, we have presented several new results on an asymptotically free,
vectorial, ${\cal N}=1$ supersymmetric gauge theory with gauge group $G$ and
$N_f$ pairs of chiral superfields in the respective representations ${\cal R}$
and $\bar {\cal R}$ of $G$, having an infrared fixed point of the
renormalization group at $\alpha_{IR}$ in the non-Abelian Coulomb phase.  At
this point, the theory has superconformal invariance.  We have derived exact
expressions for the anomalous dimension, $\gamma_{\Phi_{\rm prod}}$, of a
composite chiral superfield consisting of a (holomorphic) product of an
arbitrary number of meson, baryon, and conjugate baryon superfields $M$, $B$,
and $\tilde B$, evaluated at a superconformal IR fixed point of the
renormalization group. We have proved that $\gamma_{\Phi_{\rm prod}}$,
increases monotonically with decreasing $N_f$ in the non-Abelian Coulomb phase
of the theory and that scheme-independent expansions for these anomalous
dimensions as powers of an $N_f$-dependent variable, $\Delta_f$, exhibit
monotonic and rapid convergence to the exact $\gamma_{{\Phi}_{\rm prod}}$
throughout this phase.   However, in contrast to the behavior of
$\gamma_{M}$, which saturates its upper bound at the lower end of the NACP,
this is not, in general, the case for either $\gamma_{B}$ or $\gamma_{\Phi_{\rm
    prod}}$. In particular, $\gamma_{B}$ saturates is conformal upper bound of
1 if and only if $N_c=2$, in which case, the operator $B$ is equivalent to
$M$. Finally, we have presented and analyzed scheme-independent
calculations of the derivative of the beta function, $\beta'_{IR}$ at the
superconformal IR fixed point, up to $O(\Delta_f^3)$ for general $G$ and ${\cal
  R}$, and have given an analysis of the properties of $\beta'_{IR}$ up to 
$O(\Delta_f^4)$ for $G={\rm SU}(N_c)$ and ${\cal R}=F$.  We believe that these
new results are useful additions to the knowledge of superconformal gauge
theories.  


\begin{acknowledgments}

This research was supported in part by the Danish National
Research Foundation grant DNRF90 to CP$^3$-Origins at SDU (T.A.R.) and
by the U.S. NSF Grant NSF-PHY-16-1620628 (R.S.)

\end{acknowledgments}


\newpage


\begin{table}
  \caption{\footnotesize{
Matter content of an $\mathcal{N}=1$ supersymmetric gauge theory
    with a general complex matter representation.}}
\begin{center}
  \begin{tabular}{|c||c|c|c|c|c|}
  \hline
  & SU($N_c$) & SU($N_f$) & SU($N_f$)  & U(1) & U(1)$_R$   \\
 \hline
 $\Phi $ & ${\cal R}$  & $\fund$  & 1 & 1 & $1-[C_A/(2T_fN_f)]$ \\
  \hline
  $\tilde{\Phi}$  &  $\overline{\cal R}$ & 1 &  $\overline{\fund}$ & $-1$ & \
$1-[C_A/(2T_f N_f)]$ \\
    \hline
  \end{tabular}
  \label{sgt_table}
\end{center}
\end{table}


\begin{table}
\caption{\footnotesize{Matter content of the $\mathcal{N}=1$ 
supersymmetric gauge theory
with gauge group SU($N_c$) and $N_f$ pairs of chiral superfields in the 
fundamental and conjugate fundamental representations.}}
\begin{center}
\begin{tabular}{|c||c|c|c|c|c|}
  \hline
  & SU($N_c$) & SU($N_f$) & SU($N_f$)  & U(1)$_V$ & U(1)$_R$   \\
 \hline
 $\Phi $ & $\fund$  & $\fund$  & 1 & 1 & $1-(N_c/N_f)$ \\
  \hline
  $\tilde{\Phi}$  &  $\overline{\fund}$ & 1 &
  $\overline{\fund}$  & $-1$ & $1-(N_c/N_f)$  \\
    \hline
  \end{tabular}
\end{center}
  \label{sqcd_table}
\end{table}


\begin{table}
  \caption{\footnotesize{
Matter content of an $\mathcal{N}=1$ supersymmetric gauge theory
    with an arbitrary real or pseudoreal matter representation.}}
\begin{center}
  \begin{tabular}{|c||c|c|c|}
  \hline
  & SU($N_c$) & SU($2N_f$)  & U(1)$_R$   \\
 \hline
 $\Phi $ & ${\cal R}$  & $\fund$  & $1-[C_A/(2T_fN_f)]$ \\
 \hline
  \end{tabular}
  \label{sgtreal_table}
\end{center}
\end{table}


\begin{table}
\caption{\footnotesize{ Scheme-independent values of
$\beta'_{IR,F,\Delta_f^p}$ with $2 \le p \le 4$ for $G={\rm SU}(2)$, SU(3),
and SU(4) with $N_f$ pairs of chiral superfields in the fundamental and
conjugate fundamental representations, as functions of $N_f$, in the 
respective non-Abelian Coulomb phase intervals, $(3/2)N_c < N_f < 3N_c$.
Here, $\Delta_f=3N_c-N_f$.}}
\begin{center}
\begin{tabular}{|c|c|c|c|c|} \hline\hline
$N_c$ & $N_f$
      & $\beta'_{IR,F,\Delta_f^2}$
      & $\beta'_{IR,F,\Delta_f^3}$
      & $\beta'_{IR,F,\Delta_f^4}$
\\ \hline
2 & 3 & 1.000  & 2.167  & $-0.7482$ \\
2 & 4 & 0.444  & 0.790  & 0.214  \\
2 & 5 & 0.111  & 0.154  & 0.118  \\
\hline
3 & 5 & 0.667  & 1.296  & 0.330  \\
3 & 6 & 0.375  & 0.641  & 0.335  \\
3 & 7 & 0.167  & 0.245  & 0.185  \\
3 & 8 & 0.0417 & 0.0515 & 0.0477 \\
\hline
4 & 6 & 0.800  & 1.627  & 0.370  \\
4 & 7 & 0.555  & 1.034  & 0.428  \\
4 & 8 & 0.355  & 0.6005 & 0.352  \\
4 & 9 & 0.200  & 0.303  & 0.225  \\
4 &10 & 0.0889 & 0.1195 & 0.104  \\
4 &11 & 0.0222 & 0.02605& 0.0251 \\
\hline
\hline\hline
\end{tabular}
\end{center}
\label{betaprime_values}
\end{table}


\begin{table}
\caption{\footnotesize{ Values of $N_\ell$, $N_{f,b2z}$, $N_u$, 
$N_{f,{\rm mc},(2,3)}$, $N_{f,{\rm mc},(2,3)}-N_\ell$, 
$N_{f,{\rm mc},(3,4)}$, and $N_{f,{\rm mc},(3,4)}-N_\ell$ 
for the illustrative cases $2 \le N_c \le 4$. For notational brevity, 
the subscripts $mc$ are suppressed.}}
\begin{center}
\begin{tabular}{|c|c|c|c|c|c|c|c|} \hline\hline
$N_c$ & $N_\ell$ & $N_{f,b2z}$ & $N_u$ & $N_{f,(2,3)}$ &
$N_{f,(2,3)}-N_\ell$ & $N_{f,(3,4)}$ &
$N_{f,(2,3)}-N_\ell$
\\ \hline
2 & 3   & 3.429  & 6  & 3.428  & 0.429  & 4.799 & 1.799  \\
3 & 4.5 & 4.765  & 9  & 4.765  & 0.265  & 6.395 & 1.895  \\
4 & 6   & 6.1936 & 12 & 6.1935 & 0.1935 & 8.054 & 2.054  \\
\hline\hline
\end{tabular}
\end{center}
\label{nfvalues}
\end{table}


\begin{table}
\caption{\footnotesize{ Values of $r_\ell$, $r_{f,b2z}$, $r_u$, 
$r_{{\rm mc},(2,3)}$, $r_{{\rm mc},(2,3)}-r_\ell$, 
$r_{{\rm mc},(3,4)}$, and $r_{{\rm mc},(3,4)}-r_\ell$ in the LNN limit.}}
\begin{center}
\begin{tabular}{|c|c|c|c|c|c|c|} \hline\hline
$r_\ell$ & $r_{b2z}$ & $r_u$ & $r_{{\rm mc},(2,3)}$ &
$r_{{\rm mc},(2,3)}-r_\ell$ & $r_{{\rm mc},(3,4)}$ &
$r_{{\rm mc},(2,3)}-r_\ell$
\\ \hline
3/2 & 3/2 & 3 & 3/2 & 0 & 1.8370 & 0.3370 \\
\hline\hline
\end{tabular}
\end{center}
\label{rvalues}
\end{table}


\begin{table}
\caption{\footnotesize{ Scheme-independent values of
$\beta'_{IR,LNN,\Delta_r^p}$ with $2 \le p \le 4$ as functions of $r$ for 
$r$ in the non-Abelian Coulomb phase interval, $3/2 < r < 3$. For comparison,
we also list $\beta'_{IR,2\ell}$ (which is scheme-independent) and 
$\beta'_{IR,3\ell}$, as calculated in the $\overline{\rm DR}$ scheme. See 
text for further discussion.}}
\begin{center}
\begin{tabular}{|c|c|c|c|c|c|} \hline\hline
$r$ & 
$\beta'_{IR,2\ell}$ & 
$\beta'_{IR,3\ell}$ & 
$\beta'_{IR,LNN,\Delta_r^2}$ & 
$\beta'_{IR,LNN,\Delta_r^3}$ & 
$\beta'_{IR,LNN,\Delta_r^4}$
\\ \hline
1.5 & u      & 6.000   & 0.750  & 1.500   & 0.533   \\
1.6 & 9.800  & 3.484   & 0.653  & 1.263   & 0.529   \\
1.7 & 4.225  & 2.301   & 0.563  & 1.052   & 0.506   \\
1.8 & 2.400  & 1.604   & 0.480  & 0.864   & 0.468   \\
1.9 & 1.5125 & 1.145   & 0.403  & 0.699   & 0.419   \\
2.0 & 1.000  & 0.823   & 0.333  & 0.5556  & 0.365   \\
2.1 & 0.675  & 0.590   & 0.270  & 0.432   & 0.307   \\
2.2 & 0.457  & 0.417   & 0.213  & 0.327   & 0.249   \\
2.3 & 0.306  & 0.288   & 0.163  & 0.240   & 0.194   \\
2.4 & 0.200  & 0.193   & 0.120  & 0.168   & 0.143   \\
2.5 & 0.125  & 0.122   & 0.0833 & 0.111   & 0.0992  \\
2.6 & 0.0727 & 0.0719  & 0.0533 & 0.0676  & 0.0627  \\
2.7 & 0.0375 & 0.0373  & 0.0300 & 0.0360  & 0.03445 \\
2.8 & 0.01538& 0.01536 & 0.0133 & 0.0151  & 0.0148  \\
2.9 &0.003571&0.003570 & 0.00333& 0.003556& 0.00354 \\
3.0 & 0      & 0       & 0      & 0       & 0       \\
\hline
\hline\hline
\end{tabular}
\end{center}
\label{betaprime_values_lnn}
\end{table}


\begin{widetext}

\begin{figure}
  \begin{center}
    \includegraphics[height=6cm]{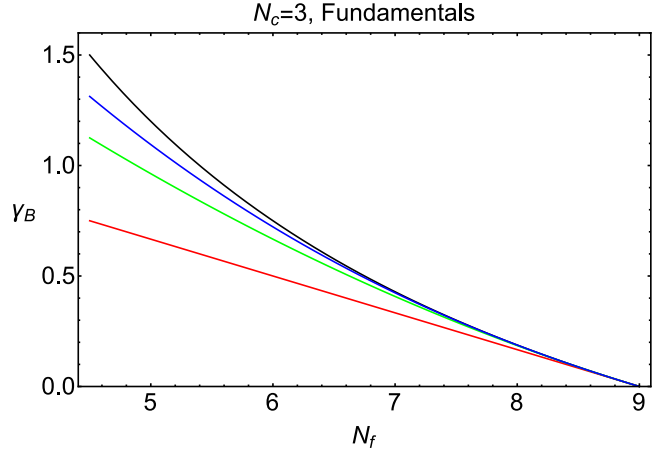}
  \end{center}
  \caption{Plot of $\gamma_{B,F,\Delta_f^p}$ = $\gamma_{\tilde B,F,\Delta_f^p}$
    with $1 \le p \le 3$, together with the exact $\gamma_{B,F}$, for $G={\rm
      SU}(3)$ and ${\cal R}=F$, as a function of $N_f$, at an IRFP in the
    non-Abelian Coulomb phase for this theory.  In this and the later figures,
    we consider $N_f$ to be generalized from integers in the NACP to real
    numbers \cite{nfintegral}. For notational simplicity, the vertical axis is
    labeled simply as $\gamma_B$. At $N_f=8$, from bottom to top, the curves
    (with colors online) refer to $\gamma_{B,F,\Delta}$ (red),
    $\gamma_{B,F,\Delta_f^2}$ (green), $\gamma_{B,F,\Delta_f^3}$ (blue), and
    the exact $\gamma_{B,F}$ (black).}
\label{baryon_Nc3_susy_fig}
\end{figure}


\begin{figure}
  \begin{center}
    \includegraphics[height=6cm]{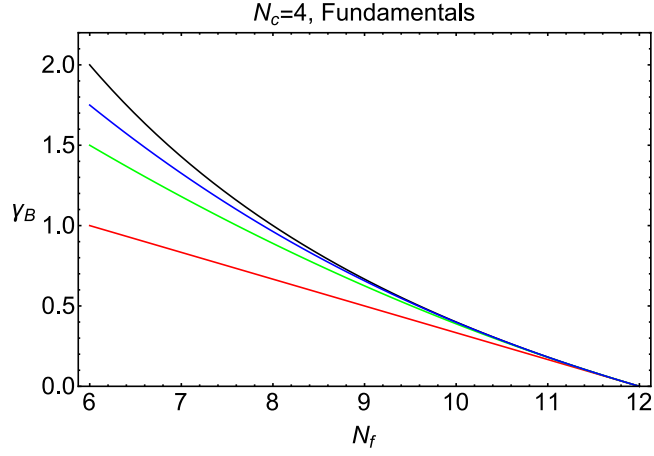}
  \end{center}
  \caption{Plot of $\gamma_{B,F,\Delta_f^p}$ = $\gamma_{\tilde B,\Delta_f^p}$ =
    with $1 \le p \le 3$, together with the exact $\gamma_{B,F}$, 
     for $G={\rm SU}(4)$ and ${\cal R}=F$, as a function of
    $N_f$, at an IRFP in the non-Abelian Coulomb phase for this theory.  
    For notational simplicity, the vertical axis is labeled
    simply as $\gamma_B$. At $N_f=8$, from bottom to top, the curves (with
    colors online) refer to $\gamma_{B,F,\Delta}$ (red),
    $\gamma_{B,F,\Delta_f^2}$ (green), $\gamma_{B,F,\Delta_f^3}$ (blue), and
    the exact $\gamma_{B,F}$ (black).}
\label{baryon_Nc4_susy_fig}
\end{figure}


\begin{figure}
  \begin{center}
    \includegraphics[height=6cm]{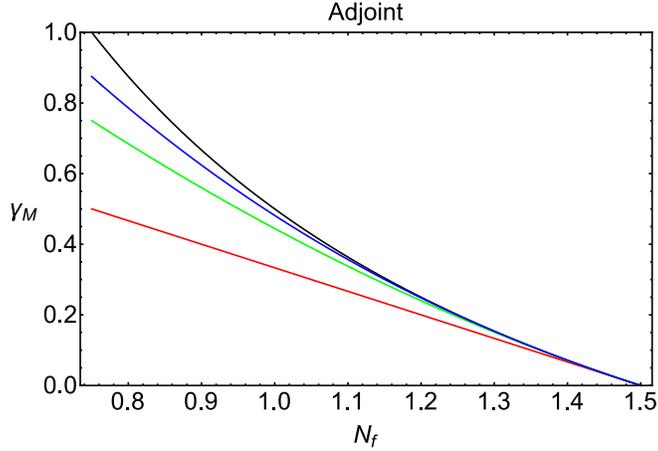}
  \end{center}
  \caption{Plot of the exact $\gamma_{M,adj}$ at an IRFP in the 
    non-Abelian Coulomb phase, together with the $O(\Delta_f^p)$
    approximations to this result with $1 \le p \le 3$, for $G={\rm SU}(4)$ and
    ${\cal R}$ equal to the adjoint representation. From bottom to top, 
    the curves (with colors online) refer to
    $\gamma_{M,adj,\Delta_f}$ (red), $\gamma_{M,adj,\Delta_f^2}$ (green), 
    $\gamma_{M,adj,\Delta_f^3}$ (blue), and the exact 
$\gamma_{M,adj}$ (black).}
\label{adjoint_susy_fig}
\end{figure}


\begin{figure}
  \begin{center}
    \includegraphics[height=6cm]{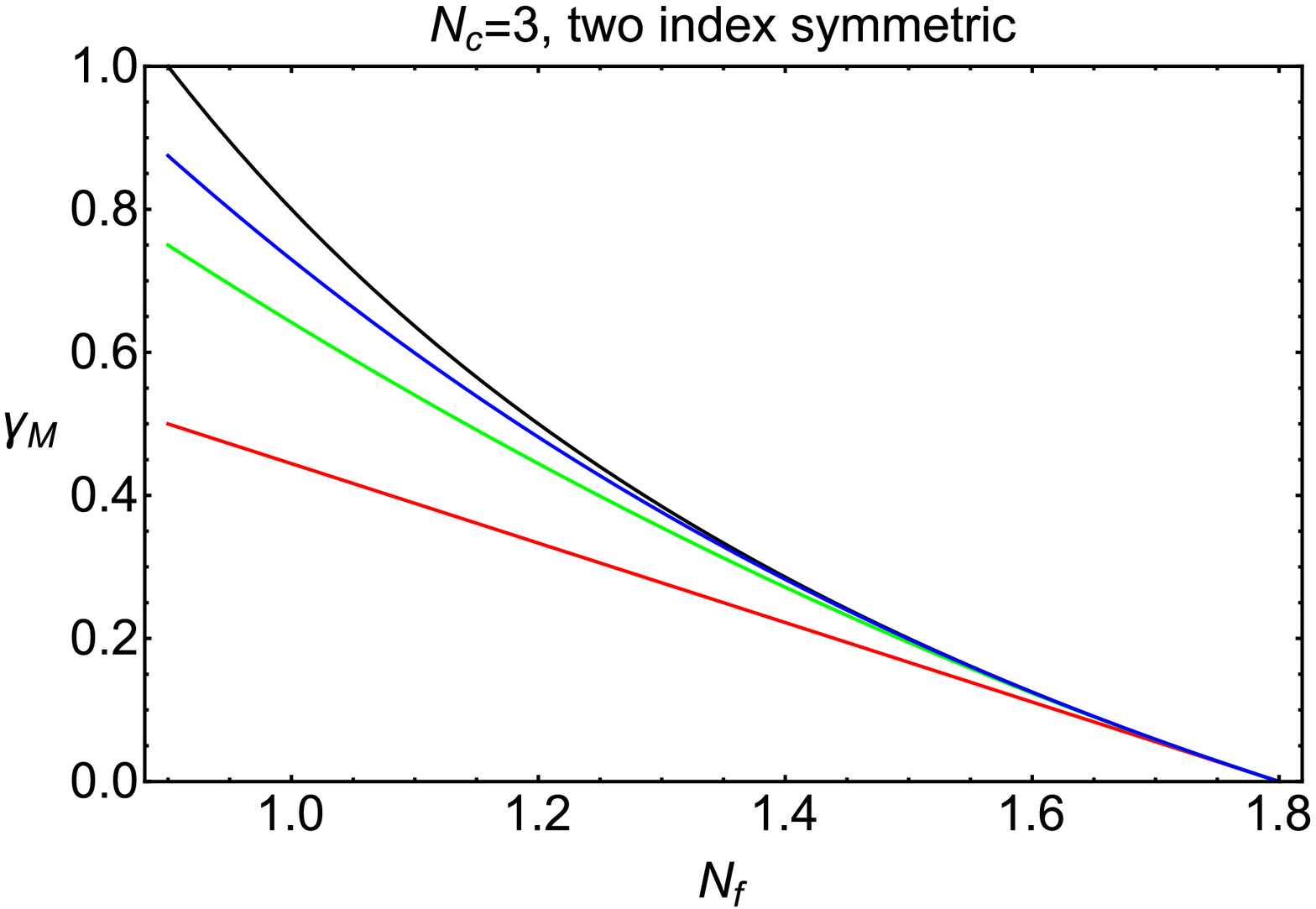}
  \end{center}
  \caption{Plot of the exact $\gamma_{M,S_2}$ at an IRFP in 
    the non-Abelian Coulomb phase, together with the $O(\Delta_f^p)$
    approximations to this result with $1 \le p \le 3$, for $G={\rm SU}(3)$ and
    ${\cal R}=S_2$, the symmetric rank-2 tensor representation. From bottom
    to top, the curves (with colors online) refer to
    $\gamma_{M,S_2,\Delta_f}$ (red), $\gamma_{M,S_2,\Delta_f^2}$ (green),
    $\gamma_{M,S_2,\Delta_f^3}$ (blue), and the exact $\gamma_{M,S_2}$
    (black).}
\label{meson_Nc3_2S_susy_fig}
\end{figure}


\begin{figure}
  \begin{center}
    \includegraphics[height=6cm]{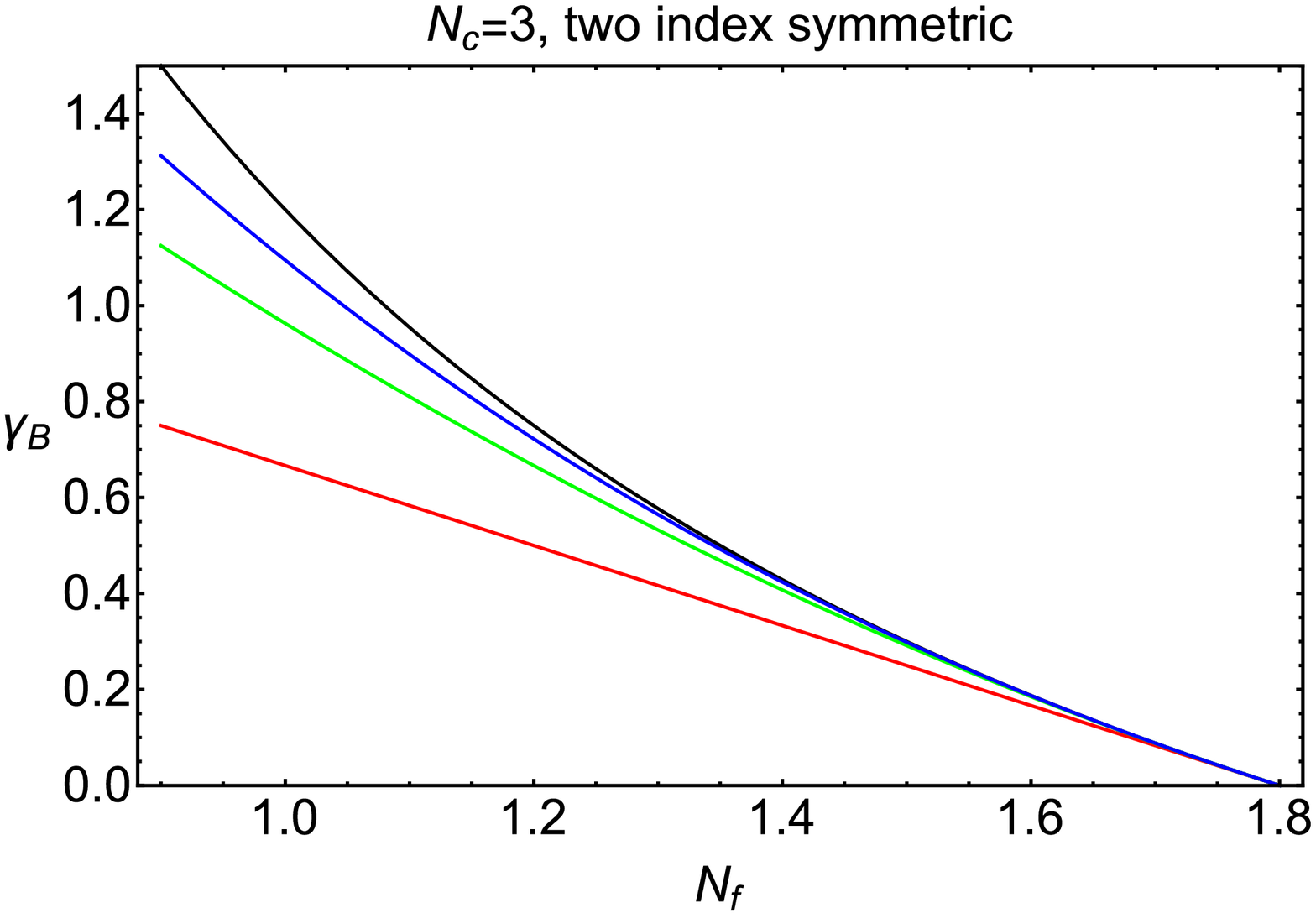}
  \end{center}
  \caption{Plot of the exact $\gamma_{B,S_2}$ at an IRFP 
    point in the non-Abelian Coulomb phase, together with the $O(\Delta_f^p)$
    approximations to this result with $1 \le p \le 3$, for $G={\rm SU}(3)$ and
    ${\cal R}=S_2$. From bottom
    to top, the curves (with colors online) refer to
    $\gamma_{B,S_2,\Delta_f}$ (red), $\gamma_{B,S_2,\Delta_f^2}$ (green),
    $\gamma_{B,S_2,\Delta_f^3}$ (blue), and the exact $\gamma_{B,S_2}$
    (black).}
\label{baryon_Nc3_2S_susy_fig}
\end{figure}


\begin{figure}
  \begin{center}
    \includegraphics[height=6cm]{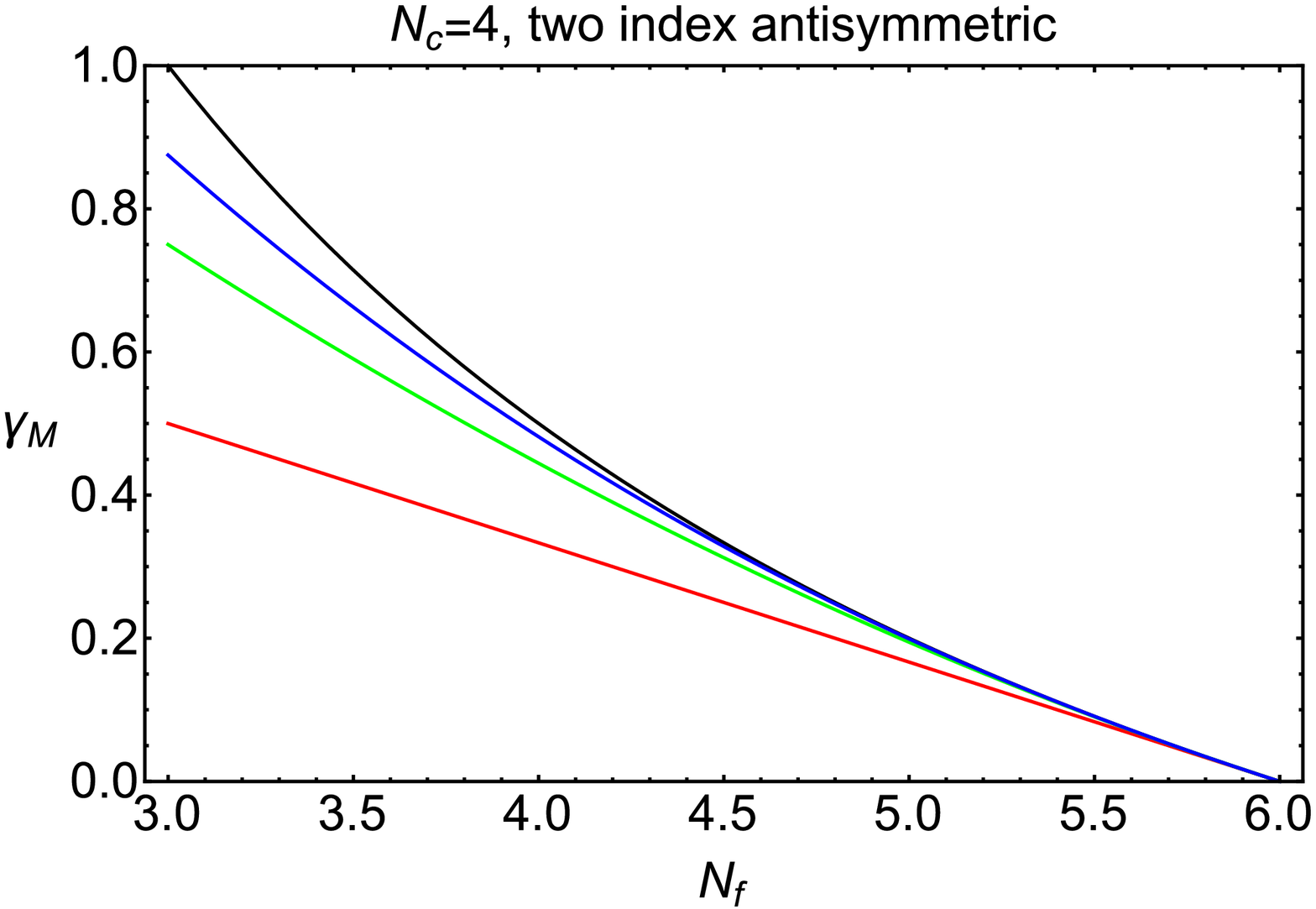}
  \end{center}
  \caption{Plot of the exact $\gamma_{M,A_2}$ at an IRFP
    in the non-Abelian Coulomb phase, together with the $O(\Delta_f^p)$
    approximations to this result with $1 \le p \le 3$, for $G={\rm SU}(4)$ and
    ${\cal R}=A_2$, the rank-2 antisymmetric tensor representation. From
    bottom to top, the curves (with colors online) refer to
    $\gamma_{M,A_2,\Delta_f}$ (red), $\gamma_{M,A_2,\Delta_f^2}$ (green),
    $\gamma_{M,A_2,\Delta_f^3}$ (blue), and the exact $\gamma_{M,A_2}$
    (black).}
\label{meson_Nc4_2A_susy_fig}
\end{figure}


\begin{figure}
  \begin{center}
    \includegraphics[height=6cm]{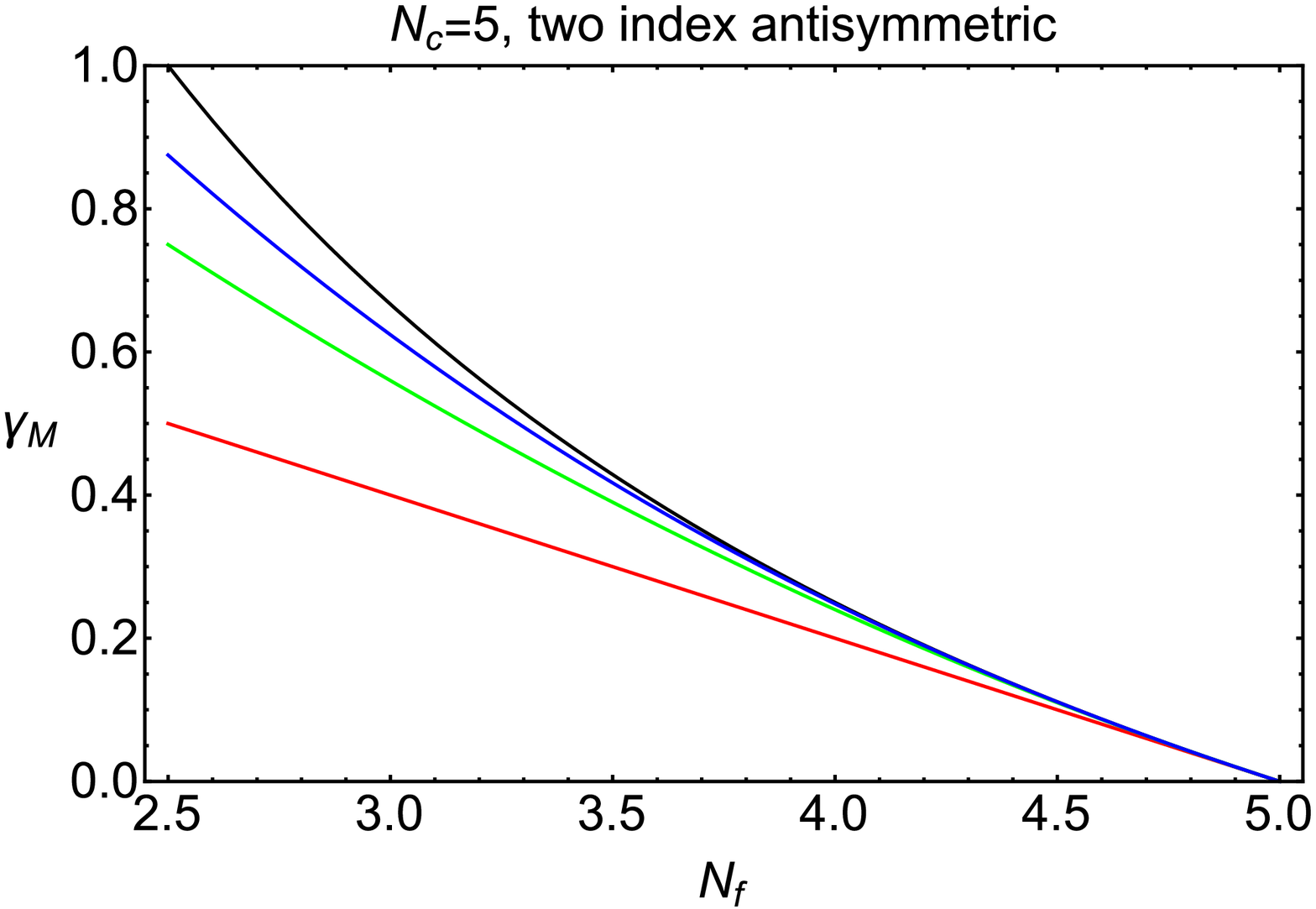}
  \end{center}
  \caption{Plot of the exact $\gamma_{M,A_2}$ at an IRFP in 
    the non-Abelian Coulomb phase, together with the $O(\Delta_f^p)$
    approximations to this result with $1 \le p \le 3$, for $G={\rm SU}(5)$ and
    ${\cal R}=A_2$. From
    bottom to top, the curves (with colors online) refer to
    $\gamma_{M,A_2,\Delta_f}$ (red), $\gamma_{M,A_2,\Delta_f^2}$ (green),
    $\gamma_{M,A_2,\Delta_f^3}$ (blue), and the exact $\gamma_{M,A_2}$
    (black).}
\label{meson_Nc5_2A_susy_fig}
\end{figure}


\begin{figure}
  \begin{center}
    \includegraphics[height=6cm]{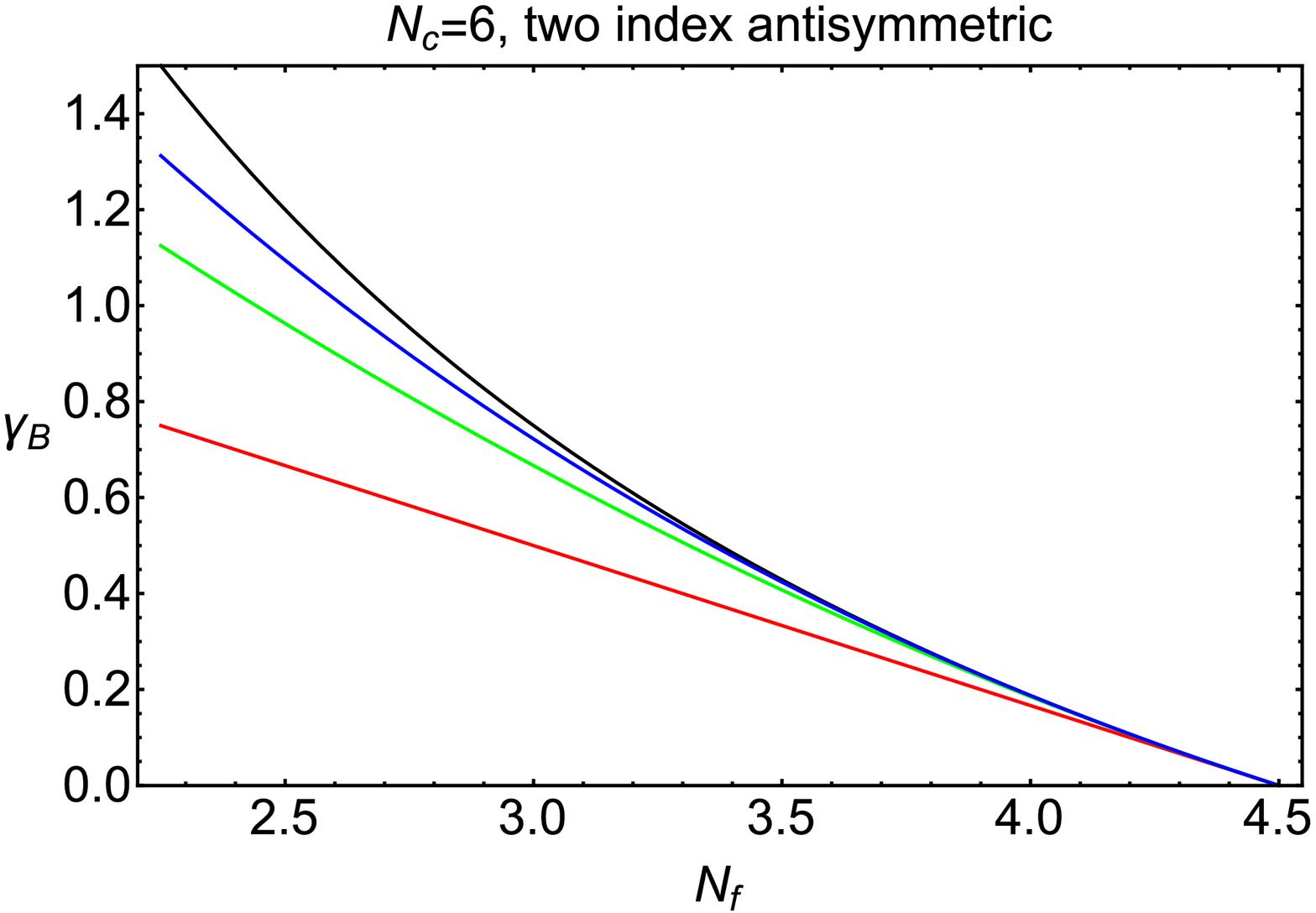}
  \end{center}
  \caption{Plot of the exact $\gamma_{B,A_2}$ at an IRFP 
    in the non-Abelian Coulomb phase, together with the $O(\Delta_f^p)$
    approximations to this result with $1 \le p \le 3$, for $G={\rm SU}(6)$ and
    ${\cal R}=A_2$. This is the special case of $\gamma_{B,A_2,Nce}$ in the
    text for $N_c=6$, where the subscript $Nce$ denotes even $N_c$. From
    bottom to top, the curves (with colors online) refer to
    $\gamma_{B,A_2,\Delta_f}$ (red), $\gamma_{B,A_2,\Delta_f^2}$ (green),
    $\gamma_{B,A_2,\Delta_f^3}$ (blue), and the exact $\gamma_{B,A_2}$
    (black).}
\label{baryon_Nc6_2A_susy_fig}
\end{figure}


\begin{figure}
  \begin{center}
    \includegraphics[height=6cm]{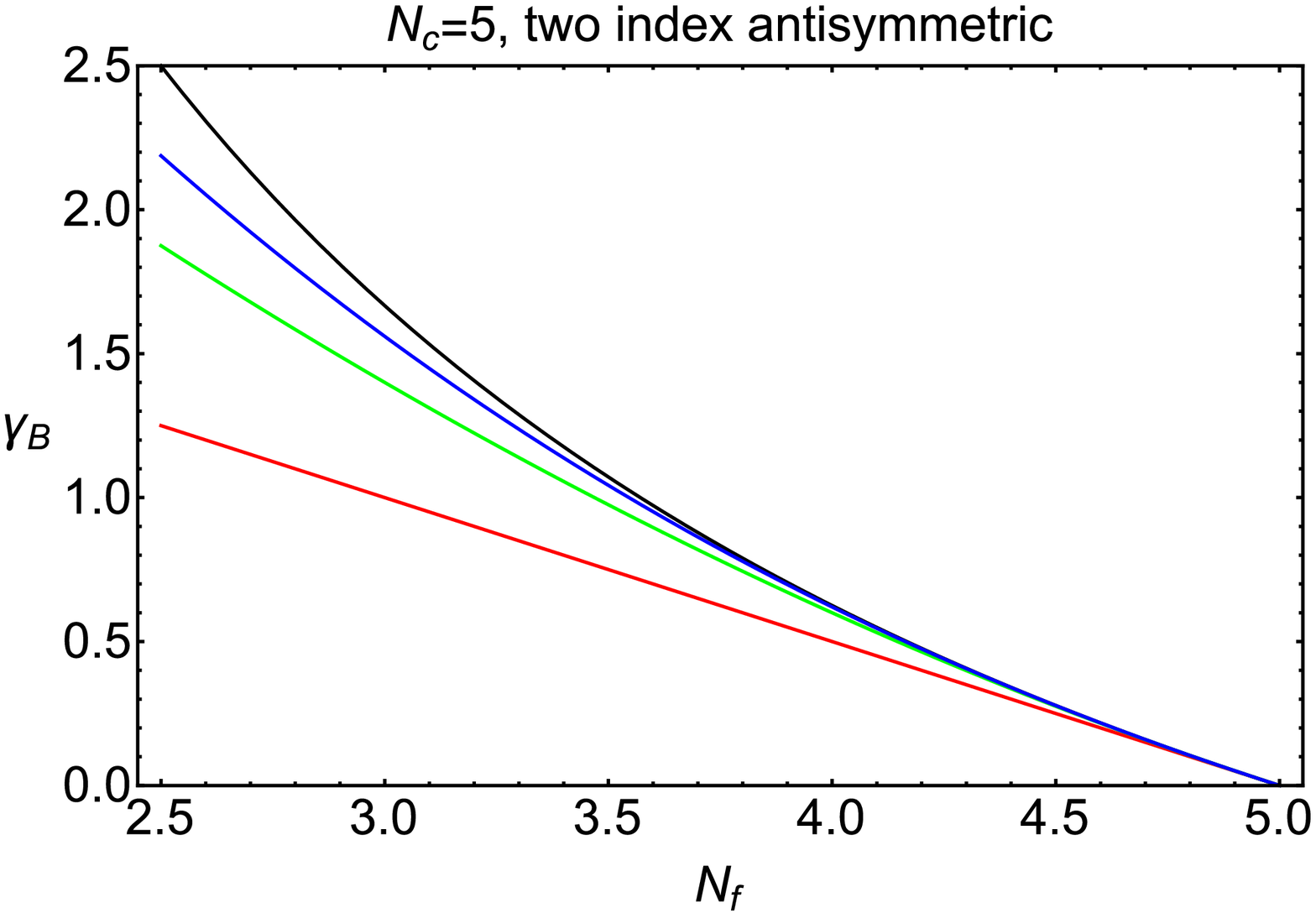}
  \end{center}
  \caption{Plot of the exact $\gamma_{B,A_2}$ at an IRFP 
    in the non-Abelian Coulomb phase, together with the $O(\Delta_f^p)$
    approximations to this result with $1 \le p \le 3$, for $G={\rm SU}(5)$ and
    ${\cal R}=A_2$. This is the special case of $\gamma_{B,A_2,Nco}$ in the
    text for $N_c=5$, where the subscript $Nco$ denotes odd $N_c$. From
    bottom to top, the curves (with colors online) refer to
    $\gamma_{B,A_2,\Delta_f}$ (red), $\gamma_{B,A_2,\Delta_f^2}$ (green),
    $\gamma_{B,A_2,\Delta_f^3}$ (blue), and the exact $\gamma_{B,A_2}$
    (black).}
\label{baryon_Nc5_2A_susy_fig}
\end{figure}


\begin{figure}
  \begin{center}
    \includegraphics[height=6cm]{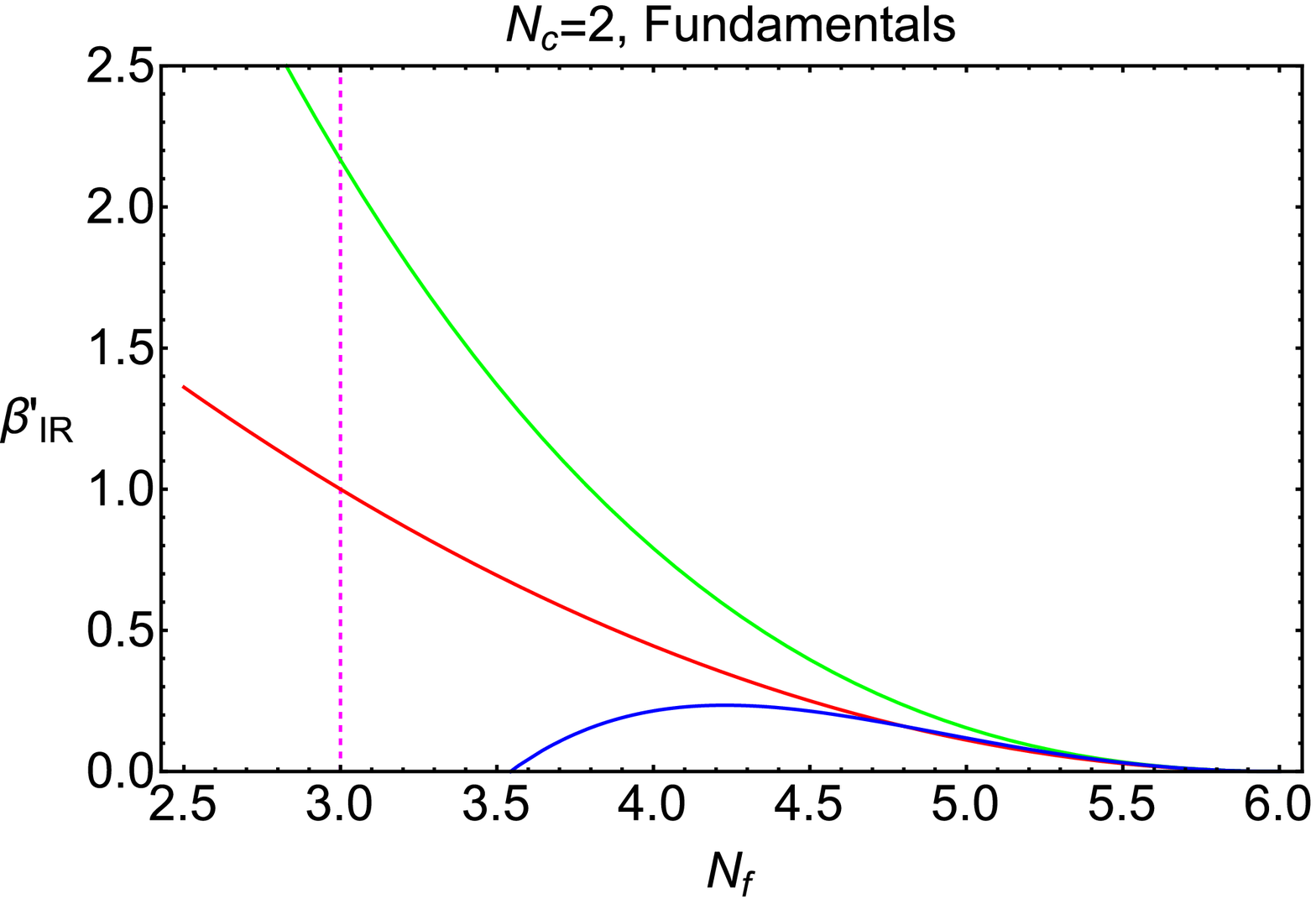}
  \end{center}
  \caption{Plot of $\beta'_{IR,\Delta_f^p}$ with $2 \le p \le 4$ for $G={\rm
      SU}(2)$ and ${\cal R}=F$, as a function of $N_f$ at an IRFP in the
    non-Abelian Coulomb phase for this theory. At $N_f=4$, from bottom to top,
    the curves (with colors online) refer to $\beta'_{IR,\Delta_f^4}$ (blue),
    $\beta'_{IR,\Delta_f^2}$ (red), and $\beta'_{IR,\Delta_f^3}$ (green).}
\label{betaprime_Nc2_susy_fig}
\end{figure}


\begin{figure}
  \begin{center}
    \includegraphics[height=6cm]{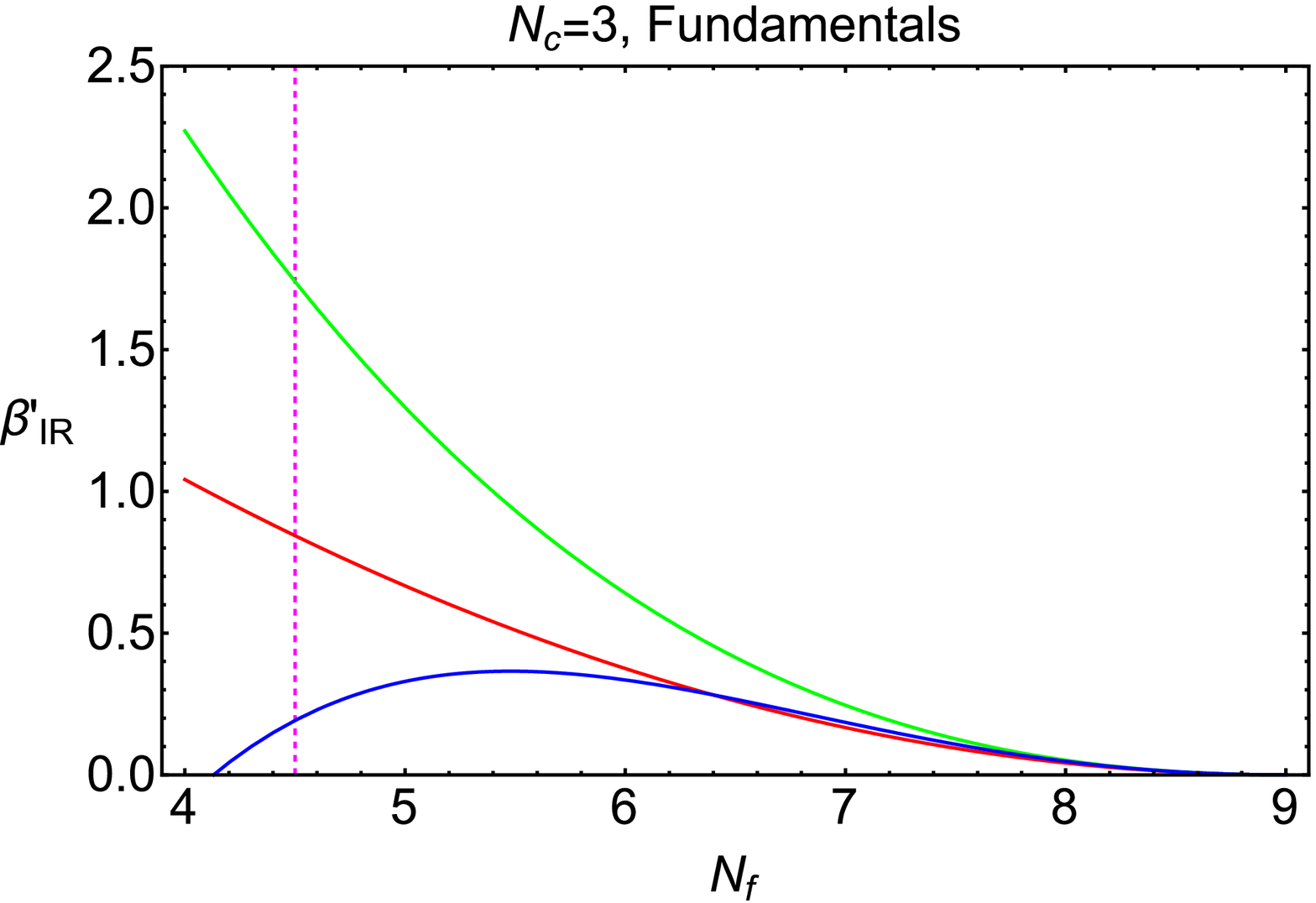}
  \end{center}
  \caption{Plot of $\beta'_{IR,\Delta_f^p}$ with $2 \le p \le 4$ for $G={\rm
      SU}(3)$ and ${\cal R}=F$, as a function of $N_f$ at an IRFP in the
    non-Abelian Coulomb phase for this theory. At $N_f=5$, from bottom to top,
    the curves (with colors online) refer to $\beta'_{IR,\Delta_f^4}$ (blue),
    $\beta'_{IR,\Delta_f^2}$ (red), and $\beta'_{IR,\Delta_f^3}$ (green).}
\label{betaprime_Nc3_susy_fig}
\end{figure}


\begin{figure}
  \begin{center}
    \includegraphics[height=6cm]{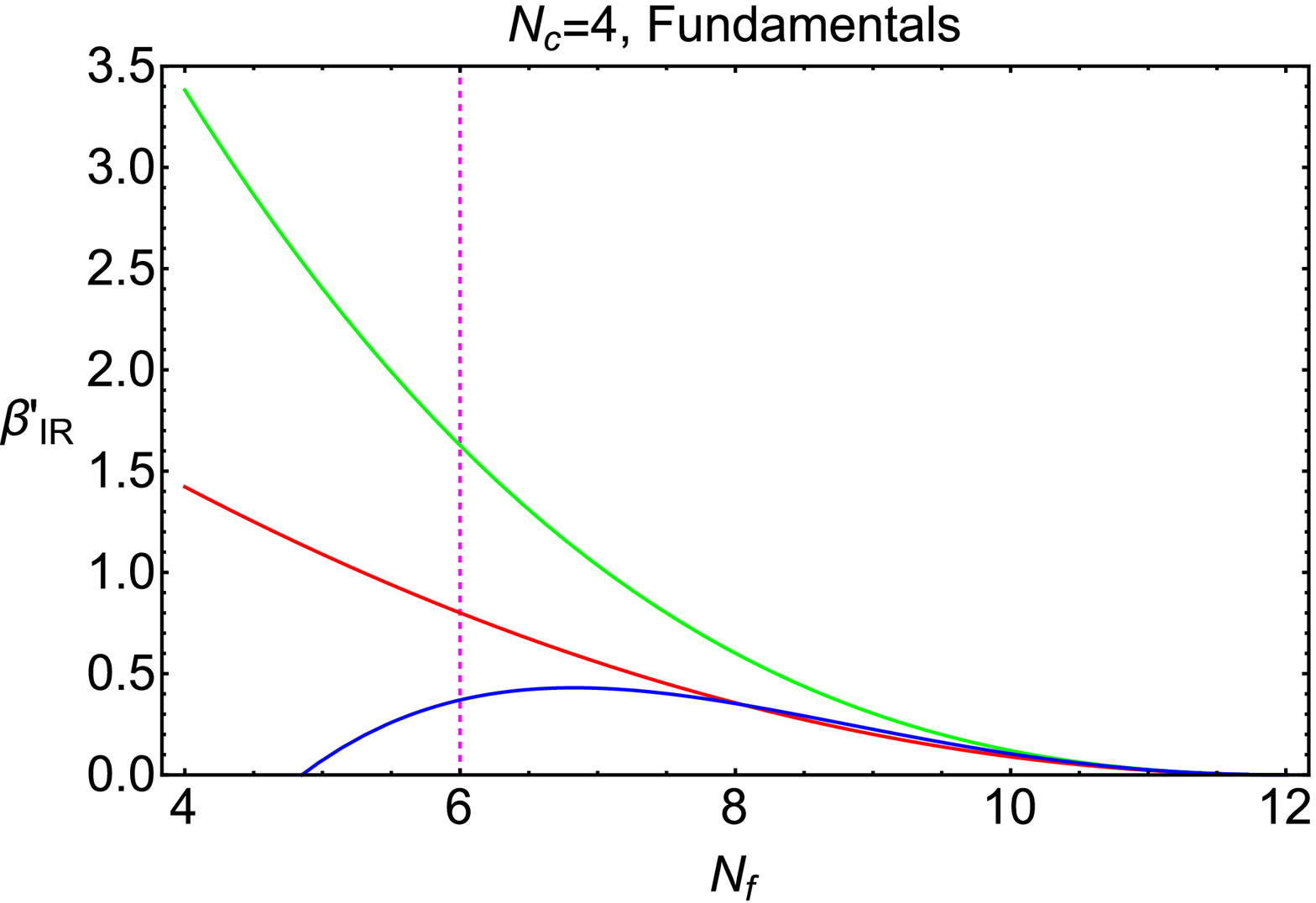}
  \end{center}
  \caption{Plot of $\beta'_{IR,\Delta_f^p}$ with $2 \le p \le 4$ for $G={\rm
      SU}(4)$ and ${\cal R}=F$, as a function of $N_f$ at an IRFP in the
    non-Abelian Coulomb phase for this theory. At $N_f=6$, from bottom to top,
    the curves (with colors online) refer to $\beta'_{IR,\Delta_f^4}$ (blue),
    $\beta'_{IR,\Delta_f^2}$ (red), and $\beta'_{IR,\Delta_f^3}$ (green).}
\label{betaprime_Nc4_susy_fig}
\end{figure}


\begin{figure}
  \begin{center}
    \includegraphics[height=6cm]{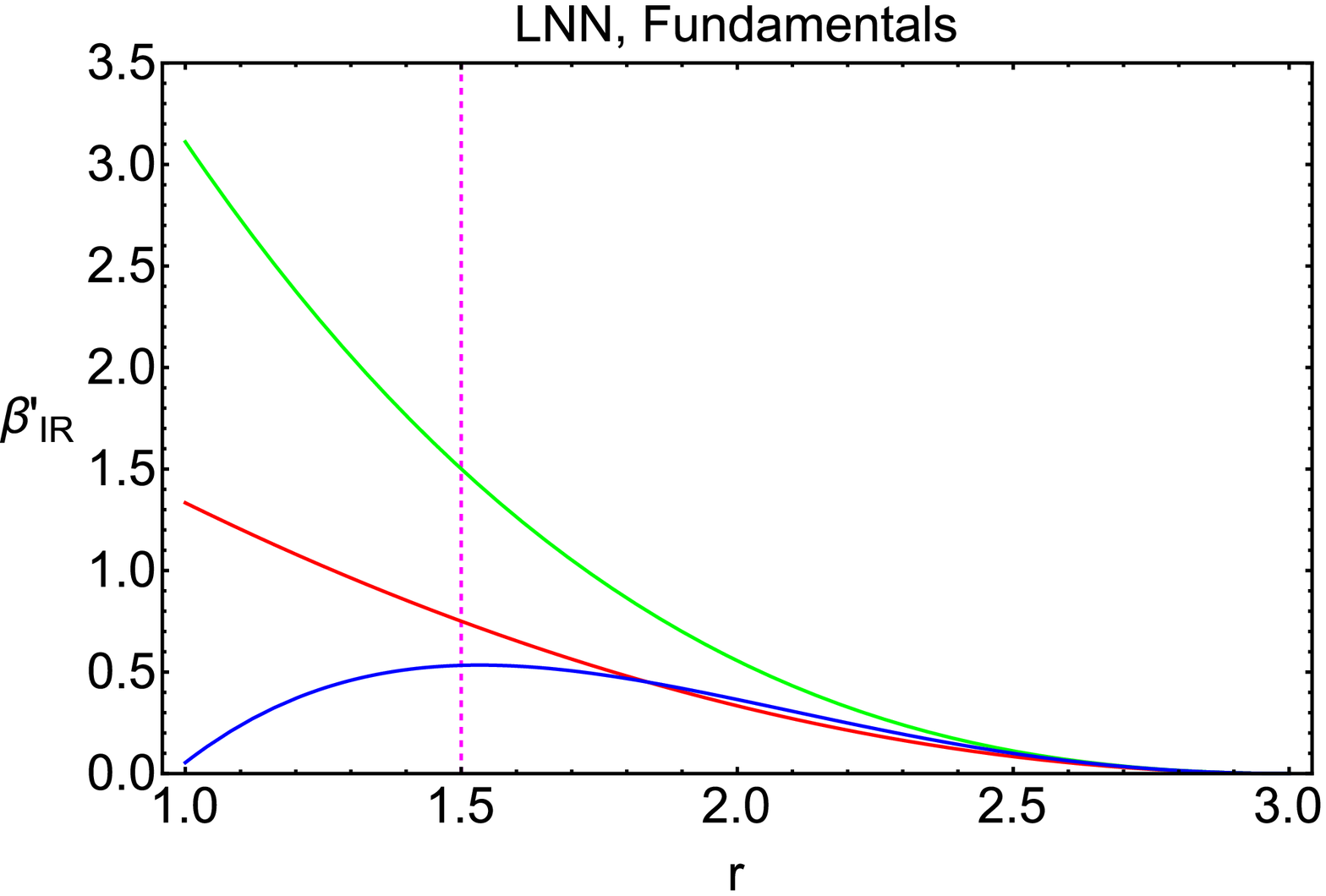}
  \end{center}
  \caption{Plot of $\beta'_{IR,LNN,\Delta_r^p}$ with $2 \le p \le 4$ as a
    function of $r$ in the LNN limit (\ref{lnn}), for $r$ at an IRFP in the
    non-Abelian Coulomb phase. At $r=1.6$, from bottom to top, the curves (with
    colors online) refer to $\beta'_{IR,LNN,\Delta_r^4}$ (blue),
    $\beta'_{IR,LNN,\Delta_r^2}$ (red), and $\beta'_{IR,LNN,\Delta_r^3}$
    (green).}
\label{betaprime_lnn_susy_fig}
\end{figure}


\end{widetext} 

\begin{thebibliography}{99}

\bibitem{fm}
%
The assumption of massless $\Phi$ incurs no loss of generality,
  since if $\Phi$ had a nonzero mass $m_0$, it would be integrated out of
  the effective field theory at scales $\mu < m_0$, and hence would not affect
  the IR limit $\mu \to 0$.

\bibitem{scalecon}
Some early studies of connections between scale and
  conformal invariance include A. Salam, Ann. Phys. (NY) {\bf 53}, 174 (1969);
  A. M. Polyakov, JETP Lett. {\bf 12}, 381 (1970);
  D. J. Gross and J. Wess, Phys. Rev. D {\bf 2}, 753 (1970); C. G. Callan,
  S. Coleman, and R. Jackiw, Ann. Phys. (NY) {\bf 59}, 42 (1970).  More recent
  works include J. Polchinski, Nucl. Phys. B {\bf 303}, 226 (1988);
  J.-F. Fortin, B. Grinstein and A. Stergiou, JHEP 01 (2013) 184 (2013);
  A. Dymarsky, Z. Komargodski, A. Schwimmer, and S. Thiessen, JHEP {\bf 10},
  171 (2015) and references therein.

\bibitem{anomdimconv}
Some authors use the opposite sign convention for the anomalous dimension,
writing $D_{\cal O} = D_{\cal O,{\rm free}} + \gamma_{\cal O}$. 

\bibitem{bz}
T. Banks and A. Zaks, Nucl. Phys. B {\bf 196}, 189 (1982).

\bibitem{gk}
E. Gardi and M. Karliner, Nucl. Phys. B {\bf 529}, 383 (1998). 

\bibitem{gg}
E. Gardi and G. Grunberg, JHEP 03, 024 (1999).

\bibitem{bfs}
T. A. Ryttov and R. Shrock, Phys. Rev. D {\bf 85}, 076009 (2012).

\bibitem{bc}
R. Shrock, Phys. Rev. D {\bf 87}, 105005 (2013).

\bibitem{lnn}
R. Shrock, Phys. Rev. D {\bf 87}, 116007 (2013). 

\bibitem{bfss}
R. Shrock, Phys. Rev. D {\bf 91}, 125039 (2015). 

\bibitem{bfs2}
G. Choi and R. Shrock, Phys. Rev. D {\bf 93}, 065013 (2016).

\bibitem{gtr}
T. A. Ryttov, Phys. Rev. Lett. {\bf 117}, 071601 (2016)
[arXiv:1604.00687]. 

\bibitem{gsi}
T. A. Ryttov and R. Shrock, Phys. Rev. D {\bf 94}, 105014 (2016).
[arXiv:1608.00068].

\bibitem{dex}
T. A. Ryttov and R. Shrock, Phys. Rev. D {\bf 94} 125005 (2016). 
[arXiv:1610.00387].

\bibitem{dexs}
T. A. Ryttov and R. Shrock, Phys. Rev. D {\bf 95}, 085012 (2017)
[arXiv:1701.06083]. 

\bibitem{dexl}
T. A. Ryttov and R. Shrock, Phys. Rev. D {\bf 95}, 105004 (2017) 
[arXiv:1703.08558]. 

\bibitem{ir}
%
Concerning notation, in earlier works in which we dealt with series
expansions for anomalous dimensions as powers of $\alpha$, we included a
subscript $IR$ when discussing the values at a conformal or superconformal
IRFP.  Here, since we will always be discussing the properties at a
superconformal theory at an IRFP, it will not be necessary to include this
subscript. Therefore, although we retain the $IR$ subscript in $\beta'_{IR}$,
we will usually omit it in the anomalous dimensions to simplify the notation.

\bibitem{gross75}
D. J. Gross, in R. Balian and J. Zinn-Justin, eds.
{\it Methods in Field Theory}, Les Houches 1975
(North Holland, Amsterdam, 1976), p. 141.

\bibitem{jones75}
D. R. T. Jones, Nucl. Phys. B {\bf 87}, 127 (1975).

\bibitem{casimir}
%
  $C_A$ and $C_f$ are the quadratic Casimir invariants for the adjoint
  representation and the fermion representation ${\cal R}$, and $T_f$ is the
  trace invariant. We use the standard normalizations for these, so that for
  $G={\rm SU}(N_c)$, $C_A=N_c^2-1$ and for ${\cal R}=F$, $C_f =
  (N_c^2-1)/(2N_c)$ and $T_f=1/2$.

\bibitem{susyloops}
M. Machacek and M. Vaughn, Nucl. Phys. B {\bf 222}, 83 (1983);
A. J. Parkes and P. C. West, Phys. Lett. B {\bf 138}, 99 (1984);
Nucl. Phys. B {\bf 256}, 340 (1985);
D. R. T. Jones and L. Mezincescu, Phys. Lett. B {\bf 136}, 242 (1984);
Phys. Lett. B {\bf 138}, 293 (1984).

\bibitem{nfintegral}
%
  Thus, if an expression for $N_f$ formally evaluates to a non-integral real
  value, it is understood implicitly that one infers an appropriate integral
  value from it.

\bibitem{nsvz}
V. A. Novikov, M. A. Shifman, A. I. Vainshtein, and V. I. Zakharov (NSVZ),
Phys. Lett. {\bf B166}, 329 (1986).

\bibitem{gracey_gammatensor}
J. A. Gracey, Phys. Lett. B {\bf 488}, 175 (2000).

\bibitem{seiberg} 
N. Seiberg, Nucl. Phys. {\bf B435}, 129 (1995).

\bibitem{susyreviews}
K. A. Intriligator and N. Seiberg, Nucl. Phys. (Proc. Suppl.) 
{\bf 45BC}, 1 (1996); 
M. A. Shifman, Prog. Part. Nucl. Phys. {\bf 39}, 1 (1997).

\bibitem{nellphysical}
%
This complication with $N_\ell$ being unphysical for odd $N_c$ is avoided in
the LNN limit (\ref{lnn}), in which one takes $N_c \to \infty$ and $N_f \to
\infty$ with the ratio $r=N_f/N_c$ fixed and finite.  In this LNN limit,
physical quantities are functions of the real variable $r$ instead of the
integer variables $N_c$ and $N_f$ and $N_\ell$ is replaced by the quantity
$r_\ell$ defined below in Eq. (\ref{rb2zdef}), which is always physical.

\bibitem{mack}
G. Mack, Commun. Math. Phys. {\bf 55}, 1 (1977)

\bibitem{dim_rcharge_rel}
M. Flato and C. Fronsdal, Lett. Math. Phys. {\bf 8}, 159 (1984); V. K. Dobrev
and V. B. Petkova, Phys. Lett. B {\bf 162}, 127 (1985).

\bibitem{susyreviews2}
See, e.g., M. F. Sohnius, Phys. Rept. {\bf 128}, 39 (1985);
J. Terning, {\it Modern Supersymmetry: Dynamics and Duality} (Oxford
University Press, Oxford, UK, 2006).

\bibitem{intriligator_wecht}
K. Intriligator and B. Wecht, Nucl. Phys. B {\bf 667}, 183 (2003) and
references therein. 

\bibitem{susyanomaly}
%
Some of the papers dealing with this issue include 
  D. R. T. Jones and J. P. Leveille, Nucl. Phys. B {\bf 206}, 473 (1982);
  D. R. T. Jones, Phys. Lett. B {\bf 123}, 45 (1983); A. I. Vainshtein,
  V. I. Zakharov, V. A. Novikov, and M. A. Shifman, JETP Lett. {\bf 40}, 920
  (1984); M. T. Grisaru and
  P. C. West, Nucl. Phys. B {\bf 254}, 249 (1985);  D. R. T. Jones,
  L. Mezincescu, and P. West, Phys. Lett. B {\bf 151}, 219 (1985); 
  M. T. Grisaru, B. Milewski, and D. Zanon, Nucl. Phys. B {\bf 266}, 589 
  (1986); P. Ensign and K. T. Mahanthappa, Phys. Rev. D {\bf 36}, 3148 (1987); 
  S. L. Adler, in
  {\it 50 Years of Yang-Mills Theories} (World Scientific, Singapore, 2005), 
  p. 187 [hep-th/0405040].

\bibitem{kataev}
A. L. Kataev and K. V. Stepanyantz, Phys. Lett. B {\bf 730}, 184 (2014);
Theor. Math. Phys. {\bf 181}, 1531 (2014).

\bibitem{gir}
B. Grinstein, K. Intriligator, and I. Rothstein, Phys. Lett. {\bf B662}, 367
(2008); for a review, see Y. Nakayama, Phys. Repts. {\bf 569}, 1 (2015).

\bibitem{minwalla}
S. Minwalla, Adv. Theor. Math. Phys. 2, 781 (1998). 

\bibitem{susyloops2}
I. Jack, D. R. T. Jones, and C. G. North, Nucl. Phys. B {\bf 486}, 479 (1996);
I. Jack, D. R. T. Jones, and A. Pickering, Phys. Lett. B {\bf 435}, 61 (1998);
A. G. M. Pickering, J. A. Gracey, and D. R. T. Jones, Phys. Lett. B {\bf 510},
347 (2001);
R. V. Harlander, D. R. T. Jones, P. Kant, L. Mihaila, and M. Steinhauser,
JHEP 0612, 024 (2006);
R. Harlander, L. Mihaila, and M. Steinhauser (HMS), Eur. Phys. J.
C {\bf 63}, 383 (2009).

\bibitem{flir}
T. A. Ryttov and R. Shrock, Phys. Rev. D {\bf 94}, 105015 (2016).

\bibitem{grisaru97}
D. Anselmi, M. T. Grisaru, and A. A. Johansen, 
Nucl. Phys. {\bf 491B}, 221 (1997). 

\bibitem{konishi}
K. Konishi, Phys. Lett. B {\bf 135}, 439 (1984).

\end{thebibliography}
\end{document}